\documentclass[aps,10pt,pre,twocolumn, superscriptaddress,longbibliography]{revtex4-2}

\usepackage[utf8]{inputenc}
\usepackage[english]{babel}
\usepackage{xcolor}
\usepackage{graphicx}
\usepackage{svg}
\usepackage[hidelinks]{hyperref}
\usepackage{amsmath}
\usepackage{amssymb}
\usepackage{amsfonts}
\usepackage{mathtools}
\usepackage{dsfont}

\usepackage{tcolorbox}
\usepackage[ruled]{algorithm2e}
\usepackage{tikz}
\usetikzlibrary{graphs, graphs.standard, quotes}
\usepackage{pgfplots}
\definecolor{bg}{rgb}{0.95,0.95,0.95}
\usepackage{hhline}
\usepackage[nameinlink]{cleveref}
\usepackage{caption}
\usepackage{subcaption}
\AddToHook{cmd/appendix/before}{%
  \crefalias{section}{appendix}%
  \crefalias{subsection}{appendix}%
}

\newcommand*\diff{\mathop{}\!\mathrm{d}}

\newcommand{\bs}[1] {\ensuremath{\boldsymbol{#1}}}

\newcommand{\prob}[1]{\mathbb{P} \left[ #1 \right]}

\newcommand{\del}{\partial}
\newcommand{\set}[2]{\left\{ #1 \right\}_{#2}}

\allowdisplaybreaks
\renewcommand{\selectlanguage}[1]{}

\begin{document}
\title{Matrix-Product Belief Propagation for continuous-state-space variables}
\date{\today}

\author{Federico Florio}
\email{federico.florio@polito.it}
\affiliation{Institute of Condensed Matter Physics and Complex Systems,
Department of Applied Science and Technology, Politecnico di Torino,
Corso Duca degli Abruzzi 24, 10129 Torino, Italy
}
\author{Alfredo Braunstein}
\email{alfredo.braunstein@polito.it}
\affiliation{Institute of Condensed Matter Physics and Complex Systems,
Department of Applied Science and Technology, Politecnico di Torino,
Corso Duca degli Abruzzi 24, 10129 Torino, Italy
}
\affiliation{Italian Institute for Genomic Medicine, Candiolo Cancer Institute,
Fondazione del Piemonte per l’Oncologia (FPO), Candiolo, 10060 Torino, Italy}
\affiliation{Istituto Nazionale di Fisica Nucleare (INFN), Via Pietro Giuria, 1, 10125 Torino, Italy}

\begin{abstract}
Computation of observables in discrete stochastic, possibly conditioned, dynamics over large sparse networks is fundamental for a myriad of applications. The Matrix-Product Belief Propagation method allows a semi-analytical estimation of such observables in tree-like networks with a controlled error that depends on the bond dimension, that is the dimension of the employed square matrices. Its computational cost is linear in the time horizon and the network size for a large family of models with discrete degrees of freedom. 
Here, a generalization of this method to models with continuous or mixed continuous/discrete degrees of freedom is presented, using a tunable expansion in a Hilbert function basis. For a wide class of models, the computational cost of the method is linear in the network size with a prefactor that depends on the basis size and the bond dimension.
The method's efficacy is demonstrated by employing a Fourier basis for a mixed continuous/discrete representation of the Kinetic Ising dynamics with real-valued random couplings, where intermediate ``local fields'' are treated as continuous. The accuracy of the method is verified via comparison with Monte Carlo simulations. For this model, we calculate time auto-correlations, time evolution of energy and magnetization, and finally we estimate the large deviation function of the magnetization at a given future time.
\end{abstract}

\maketitle

\section{Introduction}
Markov processes on networks can describe a variety of phenomena in different fields. 
We consider the problem of computing observables on dynamic trajectories, such as marginal distributions and correlations. These processes are characterized by a high-dimensional state space, which generally renders analytic treatment difficult. While in principle useful observables can be estimated numerically with controlled absolute error via Monte Carlo sampling, it is often impractically slow to achieve a reasonably small relative error. Moreover, a different approach is required when studying conditioned probabilities and rare events, as in this case dynamic trajectories must be reweighted (e.g. in an importance sampling scheme) and convergence, even to achieve a target absolute error, can become exponentially slow \cite{antulov-fantulinStatisticalInferenceFramework2014a, braunsteinInferenceConditionedDynamics2023b}. These reweighted dynamical models can be thought of as high-dimensional Hidden Markov Models and have a large range of applications, as they allow fitting an analytic description of the dynamics to empirical observations. 

Semi-analytical ``mean-field'' Statistical Physics approaches have also been devised for the estimation of marginals and observables in discrete stochastic processes on sparse networks \cite{aurellDynamicMeanfieldCavity2012,karrerMessagePassingApproach2010,braunsteinSmallcouplingDynamicCavity2025,dominguezCavityMasterEquation2020,shresthaMessagepassingApproachRecurrentstate2015,lokhovInferringOriginEpidemic2014}. Among these approaches, only \cite{karrerMessagePassingApproach2010} is asymptotically exact in acyclic interaction graphs, but it cannot be applied to the reweighted case. The Dynamic Belief Propagation (DBP) \cite{altarelliLargeDeviationsCascade2013,altarelliBayesianInferenceEpidemics2014,altarelliLargeDeviationsCascade2013,muntoniEffectivenessProbabilisticContact2024,braunsteinStatisticalMechanicsInference2023} method has been devised to treat the reweighted case  and is asymptotically exact on locally tree-like graph ensembles. While being general in principle, in practice DBP is effective only on a certain subclass of \textit{non-recurrent} models, characterized by a polynomial number of single-site trajectories in the time horizon and becomes exponentially slow otherwise. Note that \cite{karrerMessagePassingApproach2010} is also restricted to non-recurrent models. Recently, the Matrix Product Edge Messages method \cite{barthelMatrixProductAlgorithm2018,barthelMatrixProductApproximation2020a}, based on Dynamic Cavity \cite{neriCavityApproachParallel2009}, has been introduced to overcome this limitation, and has later been extended to a DBP scheme called Matrix-Product Belief Propagation (MPBP) \cite{crottiMatrixProductBelief2023c}. MPBP employs the matrix-product-state approximation to obtain linear computational complexity in the time horizon even for many recurrent models with a potentially exponential number of single-site trajectories, and can naturally treat the reweighted case. MPBP, however, is, like most of these methods, restricted to discrete state variables. Moreover, an efficient calculation of the update in the approach relies on a recursive computation involving discrete intermediate variables, which further limits its applicability.
Here, a generalization of MPBP to account for continuous state variables is developed, based on an expansion in a Hilbert function basis and therefore called Basis-MPBP.
The approximation is tunable and can be made arbitrarily precise by increasing the number of basis elements and corresponding parameters.\\
Reweighted dynamics in continuous or mixed continuous-discrete space are sometimes called State-Space models and have several applications besides the one presented in this work \cite{auger-metheGuideStateSpace2021,pattersonStateSpaceModels2008,catorStateSpaceAgentBasedModel2026,rasmussenInferenceNonlinearEpidemiological2011}. While the method can in principle be applied to mixed continuous-discrete space, the application shown in this work is on the Kinetic Ising dynamics with real-valued random couplings. Although it is a discrete-variable model, the naturally arising intermediate variables are the local fields, which are essentially continuous. In that situation, the efficient calculation in the MPBP update step is still not possible, making it unfeasible even for moderately large degrees. With Basis-MPBP, the computational cost becomes linear in the degree, allowing to study several classes of graphs that were intractable before. In the case of the Kinetic Ising dynamics, a natural choice is the Fourier basis, leading to a scheme in which intermediate quantities correspond to frequency probability distributions.\\

\Cref{sec:models-methods} will describe the methods and the models they can be applied to, and \cref{sec:glauber} will show the specific treatment for the main application of this work, which is the Kinetic Ising dynamics with disordered couplings. Finally, results are shown in \cref{sec:results}.

\section{Models and methods}
\label{sec:models-methods}
Given a graph $\mathcal{G}=(\mathcal{V},\mathcal{E})$, with $\mathcal{V}= \{1, \dots, N\}$ and $\mathcal{E} \subseteq \mathcal{V}\times\mathcal{V}$, call $x_i^t$ the state of node $i$ at the (discrete) time instant $t$. Consider the following probability density function for discrete-time trajectories $\overline{x}_i = (x_i^0, x_i^1, \dots, x_i^T)$ 
\begin{equation}
    p(\overline{\bs{x}}) = \frac{1}{Z} \prod_{i=1}^N f_i^0(x_i^0) \prod_{t=1}^{T} f_i^t(x_i^t, x_i^{t-1}, \bs{x}_{\del i}^{t-1})\label{eq:dynamics}
\end{equation}
where the family of functions $f_i^t(x_i^t, x_i^{t-1}, \bs{x}_{\del i}^{t-1})$ take real non-negative values and $\del i$ indicates the neighbors of node $i$. 
This formalism is very general and can describe most (possibly reweighted) parallel Markov dynamics on networks. For normal (non-reweighted) Markov dynamics, the $f_i^t$ terms are simply the transition weight, $f_i^t(x_i^t, x_i^{t-1}, \bs{x}_{\del i}^{t-1})=\mathbb{P}[x_i^t | x_i^{t-1}, \bs{x}_{\del i}^{t-1}]$ and $Z=1$. However, analytical computations on \cref{eq:dynamics} such as the computation of averages over trajectories are typically very difficult, as it involves integrating over the $N T$ variables $\{x_i^t\}$.\\
Matrix-Product Belief Propagation \cite{crottiMatrixProductBelief2023c} relies on the following factor graph \cite{mezardInformationPhysicsComputation2009}: each original vertex becomes a factor vertex in the factor graph and each original edge is ``split" into two by a new variable vertex which carries the joint trajectories of the two endpoints (see \cite{altarelliLargeDeviationsCascade2013}, fig. 3). The BP equations for messages $\mu_{ij}(\overline{x}_i, \overline{x}_j)$ can be written:
\begin{equation}\label{eq:bp-update-step}
\begin{split}
    \mu_{ij}(\overline{x}_i, \overline{x}_j) = \int \prod_{k\in\del i\setminus j} \prod_{t=0}^{T} \diff{x_k^t} f_i^0(x_i^0) &\\ \prod_{t=1}^{T} f_i^t(x_i^t, x_i^{t-1}, \bs{x}_{\del i}^{t-1})
    &\prod_{k\in\del i\setminus j} \mu_{ki}(\overline{x}_k, \overline{x}_i).
\end{split}
\end{equation}
Following \cite{crottiMatrixProductBelief2023c}, each message is represented as a matrix-product state:
\begin{equation}\label{eq:mps-ansatz}
    \mu_{ij}(\overline{x}_i, \overline{x}_j) = A_{ij}^0(x_i^0, x_j^0) A_{ij}^1(x_i^1, x_j^1) \dots A_{ij}^T(x_i^T, x_j^T)
\end{equation}
with $A_{ij}^t(x_i^t, x_j^t) : \mathbb{R}^2 \to \mathbb{R}^{d_t,d_{t+1}}$. Plugging the matrix-product \textit{ansatz} (\ref{eq:mps-ansatz}) into the BP update equations (\ref{eq:bp-update-step}) one obtains:
\begin{equation}
    \mu_{ij}(\overline{x}_i, \overline{x}_j) = \prod_{t=0}^T B_{ij}^t(x_i^{t+1}, x_i^t, x_j^t)
\end{equation}
\begin{equation}
\begin{split}
    B_{ij}^t(x_i^{t+1}, x_i^t, x_j^t) = \int \prod_{k\in\del i\setminus j} &\diff{x_k^t} f_i^{t+1}(x_i^{t+1}, x_i^t, \bs{x}_{\del i}^t) \\
    &\bigotimes_{k\in\del i\setminus j} A_{ki}^t(x_k^t, x_i^t).
\end{split} \label{eq:bp-update-matrix}
\end{equation}
This is not yet in the form of \cref{eq:mps-ansatz}. For discrete variables $x$ as in the conditions of \cite{crottiMatrixProductBelief2023c}, one can separate the dependence of $B_{ij}^t(x_i^{t+1}, x_i^t, x_j^t)$ on different times. Indeed, it can always be written
\begin{align*}
    \left[ B_{ij}^t(x_i^{t+1}, x_i^t, x_j^t) \right]_{a,b} =& \sum_{y,c} \left[ B_{ij}^t(y, x_i^t, x_j^t) \right]_{a,c} \delta_{y, x_i^{t+1}} \delta_{c,b} \\
    =& \sum_{y,c} \left[ C^t(x_i^t, x_j^t) \right]_{a, (y,c)} \\
    &\left[ D^t(x_i^{t+1}) \right]_{(y,c), b}
\end{align*}
and one can close the matrix-product \textit{ansatz} by recovering the updated matrices $A_{ij}^t(x_i^t, x_j^t) = D^{t-1}(x_i^t) C^t(x_i^t, x_j^t)$. This is possible because one can freely exchange ``physical" variables and matrix indices. When variables are continuous, this can no longer be done and one has to do some additional work.\\
Moreover, notice that, to perform the BP update equations numerically, some parametrization should be assumed for the functional forms of the matrices $A_{ki}^t$ and $B_{ij}^t$ of \cref{eq:bp-update-matrix}.
In the following, Basis-MPBP is introduced, which is a method based on an expansion in a basis of functions and solves both of these problems. When expanding in a basis, integrals can be performed easily and the equations can be closed under the matrix-product \textit{ansatz} as shown in the next sections.

\subsection{Basis choice and expansion}
\label{sec:basis-expansion}
Taking $\{u_\alpha\}, u_\alpha: I \to \mathbb{C}$ a basis of orthonormal functions on an interval $I\subseteq \mathbb{R}$, the matrices $A_{ij}^t(x_i^t, x_j^t)$ can be written as
\begin{equation}\label{eq:basis-expansion}
    A_{ij}^t(x_i^t, x_j^t) = \sum_{\alpha, \beta} A_{ij}^t[\alpha, \beta] u_\alpha(x_i^t) u_\beta(x_j^t)
\end{equation}
where the entries of the matrices $A_{ij}^t[\alpha, \beta]$ are the coefficients of the expansions of the entries of $A_{ij}^t(x_i^t, x_j^t)$ in the product basis of $\{u_\alpha\}$, that is to say
\begin{equation}
\begin{split}
    \left[A_{ij}^t[\alpha, \beta]\right]_{m,n} = \int \diff{x_i^t}\diff{x_j^t} &\left[A_{ij}^t(x_i^t, x_j^t)\right]_{m,n}\\
    &u_\alpha^*(x_i^t) u_\beta^*(x_j^t)    
\end{split}
\end{equation}
The functional dependence on $x_i^t$ and $x_j^t$ has been expressed by means of the basis.  A truncation of the infinite series leads to an approximation with a locally controlled error in $L^2$ norm thanks to Parseval's theorem.\\
By plugging the expansion in \cref{eq:basis-expansion} into \cref{eq:bp-update-matrix}, one finds that the BP equations are closed under the \textit{ansatz}:
\begin{align}
    \mu_{ij} \left( \overline{x}_{i}, \overline{x}_{j} \right) = \prod_{t} \sum_{\delta_{i}^{t}, \gamma_{j}^{t} } C_{ij}^{t} \left[ \delta_{i}^{t}, \gamma_{j}^{t} \right] u_{\delta_{i}^{t}} \left( x_{i}^{t} \right) u_{\gamma_{j}^{t}} \left( x_{j}^{t} \right).
\end{align}
When doing this, one needs to follow the steps below
\begin{enumerate}
    \item The transition weight $f_i^t(x_i^t,x_i^{t-1},x_{\del i}^{t-1})$ has to be projected onto the basis (this can be done only once, before iterating).\label{item:a}
    \item The Kronecker product of the matrices of input messages must be projected onto the basis.\label{item:b}
    \item The resulting expression for the output message has a form similar to \cref{eq:basis-expansion}, but with matrices that depend on consecutive times. An SVD procedure is employed to disentangle it into the form of \cref{eq:basis-expansion}.\label{item:c}
    \item When performing the BP updates, the matrices grow in size. To avoid this, a second SVD procedure must be employed, truncating the matrices to the desired size. Similarly to the discrete case, this gives a (locally) controlled approximation (details in \cref{app:bound-SVD}). \label{item:d}
\end{enumerate}

The detailed calculation is reported in \cref{app:mpbp-basis}. Regarding the choice of the basis, two considerations are important. First, calculations in points \ref{item:a}-\ref{item:c} must be performed efficiently. Ideally, one seeks  a closed analytical expression or a very fast numerical estimation (with the possible exception of point \ref{item:a} for which a slower calculation could be employed since the computation must be done only once). Additionally, some bases could be preferred because they represent belief propagation messages for a given model more naturally, e.g. with smaller residual error at a fixed number of basis elements. In this regard, for some choices of the basis, one could in principle evaluate the square norm of the functions being approximated and compare it with the sum of the squares of the coefficients of the approximant, to obtain the residual error of the approximation. Similarly, when truncating discrete matrix-product states, it is possible to control the error, as explained in \cite{oseledetsTensorTrainDecomposition2011a}; the resulting error, as proved in \cref{app:bound-SVD}, bounds the Frobenius norm of the difference between the original function and the truncated matrix-product state expressed in the basis.\\
While in principle one could define a target approximation accuracy and increase the bond dimension $d$ and the basis size $K$ for every single approximation that is performed during iteration until they meet such accuracy, this strategy is computationally expensive and computes values with needlessly high precision even in the first iterations, in which messages are still far away from the fixed point. In practice, we employed the following general ``refining'' strategy: starting from initial values of the parameters $d$ and $K$, iterate Basis-MPBP until a quasi-fixed point of the messages is found. Then increase the parameters and continue iterating starting from the previously found messages, repeating this procedure until the results become stable (up to a target precision). 

\begin{figure*}[ht]
    \centering
    \includegraphics[width=0.8\linewidth]{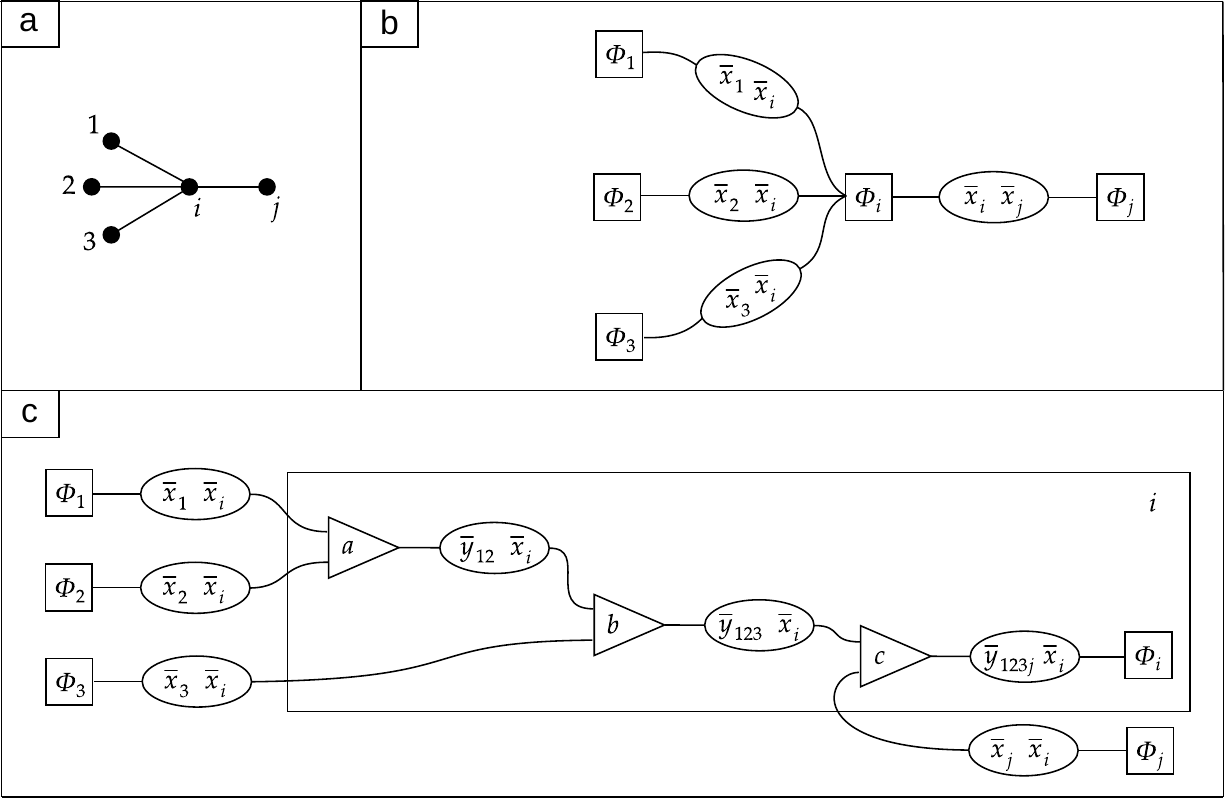}
    \caption{Representation of how a factor graph can be redesigned to deal with high-degree nodes. Panel a shows the original interaction network; panel b shows the dual representation of the factor graph (see \cite{altarelliLargeDeviationsCascade2013}, fig. 3). In panel c, the factor corresponding to node $i$ is modified: the new construction uses only two types of simpler factors to perform all the operations that took place in the previous formulation.}
    \label{fig:alternative-factor-graph}
\end{figure*}

\begin{figure*}[ht]
    \hspace{3cm}
    \includegraphics[width=0.8\linewidth]{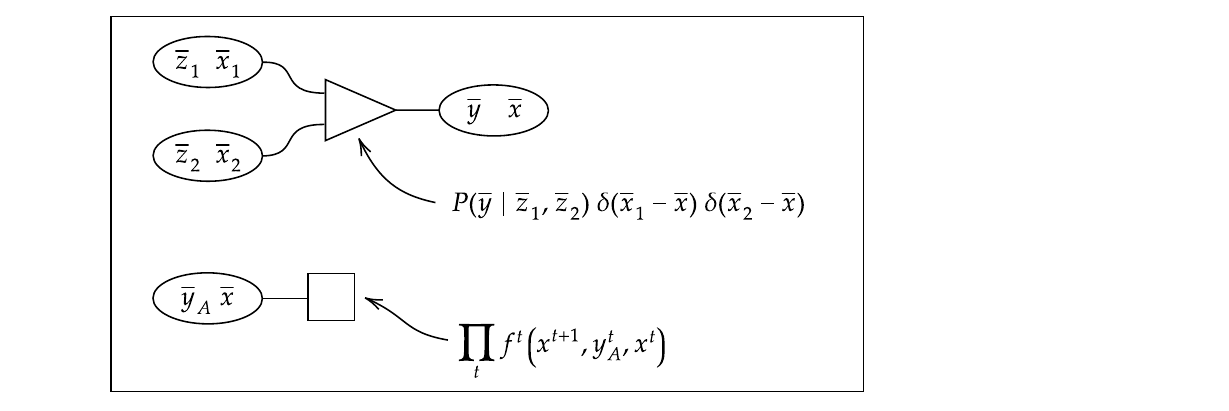}
    \caption{Two factors are needed: the triangular ones enforce the transformations from the original variables $\overline{z}$ to the new auxiliary variables $\overline{y}$, the square ones provide the dynamical update.}
    \label{fig:factors}
\end{figure*}

\subsection{Efficient calculation of MPBP equations for large degrees}
\label{sec:mpbp-alternative}
An iterative procedure was proposed in \cite{crottiMatrixProductBelief2023c} to compute the BP updates \cref{eq:bp-update-step} in a time that is linear in the degree $c$ for a subclass of models. An equivalent alternative formulation is presented here that just involves a modified factor graph featuring $\mathcal{O}(c)$ simpler factors, each with degree at most 3. 

As illustrated in \cref{fig:alternative-factor-graph}, the modified factor graph decomposes the iterative calculation into several easier computations, represented by factors having degree 1 or 3.
Some of the factors, depicted as triangles, enforce the constraints between the original and auxiliary variables, while others, represented as squares, are used to encode the dynamical process. A representation of the two types of factors can be found in \cref{fig:factors}. In order to calculate the outgoing messages towards the $\Phi$ factors, one can proceed in two consecutive sweeps: first, messages from left to right are computed (from $a$ to $b$, then from $b$ to $c$ and so on); subsequently, messages from right to left are computed (from $i$ to $c$ and so on). This whole procedure gives the same result as the standard BP iteration, as shown in the detailed calculation reported in \cref{app:alternative-factor-graph-calculation}, which shows how to compute the outgoing messages from node $i$ in the network depicted in \cref{fig:alternative-factor-graph}.

This formulation also clarifies the class of models for which such efficient updates are possible. All models for which the constraints of the dynamics can be recast in this new factor graph have an update which is linear in the degree. The reason is simply that the update is linear in the number of factors in the modified factor graph, and the number of new factors is linear in the degree of the old graph. In the new factor graph, all factors have degree at most 3, so the dependence on the degree becomes a constant factor. Models for which the interaction of each node with its neighbors is mediated by a single scalar quantity which is associatively computed from the neighbors' states (as the ones mentioned in \cite{crottiMatrixProductBelief2023c}) clearly fall into this class.\\
This formulation is also useful because it makes it possible to deal with cases of mixed nature, in which, for example, the dynamical variables $\overline{x}_i$ are discrete, but there is a need for continuous intermediate variables $\overline{y}_A$. When this occurs, the transformations between discrete and continuous values are performed by the triangular factors. If the original model is fully continuous, the triangular factors simply do not include this conversion.

\section{Application to Kinetic Ising}
\label{sec:glauber}
Basis-MPBP is applied to the Kinetic Ising model, in which the variables are Ising spins $x_i^t\in\{\pm1\}$ and which stochastically update their values in time according to the following transition probability
\begin{equation}
\begin{split}
    \prob{\bs{x}^{t+1} | \bs{x}^t} =& \prod_{i=1}^N \left\{\rho\delta_{x_i^{t+1},x_i^t}+(1-\rho) \vphantom{\frac{\exp\left[{ h_i + \sum_{j\in\del i}}\right]}{\cosh \sum_{j\in\del i}}} \right. \\
    & \left. \frac{\exp\left[{\beta \left( h_i + \sum_{j\in\del i} J_{ji} x_j^t \right) x_i^{t+1}}\right]}{2\cosh\left[\beta \left( h_i + \sum_{j\in\del i} J_{ji} x_j^t \right)\right]} \right\}\label{eq:kineticising}
\end{split}
\end{equation}
where $\rho$ controls an ``inertial'' term.
In the $\rho\to 1$ limit, the update converges to that of the sequential \textit{Glauber} dynamics in continuous time, and for $\rho=0$ the update corresponds to the standard Kinetic Ising model.
In this system, the variables are of course discrete, but when the couplings $J_{ij}$ take disordered real values, computations for nodes of even moderate degree become impractical.
Indeed, as explained in \cite{crottiMatrixProductBelief2023c}, for a class of models including Kinetic Ising, computations of MPBP can be performed linearly in the time horizon by resorting to intermediate variables $y_A^t,\;\; A\subseteq\del i$ to express the functional dependence of $\prob{x_i^{t+1} | x_i^t, \bs{x}_{\del i}^t}$ on $\bs{x}_{\del i}^t$. In the particular case of Kinetic Ising, those variables are the local fields due to the neighbors of node $i$: $y_A^t = \sum_{j\in A} J_{ji} x_j^t$, and, in the case of disordered couplings, the domains of such variables grow exponentially with $|A|$ (on the contrary, in the case of equal couplings, the domains would grow linearly with $|A|$). To avoid this exponential growth, the $y_A^t$ variables can be considered as continuous and the formulation introduced in \cref{sec:mpbp-alternative} can be used to perform the computations.\\
\Cref{alg:glauber-mpbp-update} illustrates how outgoing messages from a node are computed, and in the following sections a detailed description of each step is provided. The Fourier basis $u_\alpha(x) = e^{\frac{2\pi i}{P} \alpha x}$ will be employed, together with the scalar product $\langle f | g \rangle \coloneqq \frac{1}{P} \int_{-P/2}^{+P/2} f^*(x) g(x) \diff{x}$. This choice was dictated by the first consideration in \cref{sec:basis-expansion}. As the transition weight $f_i^t$ depends on the local field $\sum_{j\in \del i} J_{ji} x_j^{t-1}$, the resulting operation is a convolution, which is particularly convenient in the Fourier basis. The basis is truncated to a finite number of elements, so that a function is approximated by $f(x) \simeq \sum_{\alpha=-K}^K f[\alpha] u_\alpha(x)$.

\begin{tcolorbox}[float*=htbp, width=\textwidth]
\setlength{\algoheightrule}{0pt}
\setlength{\algotitleheightrule}{0pt}
\begin{algorithm}[H]
    \caption{Computation of outgoing messages from node $i$\\}
    \label{alg:glauber-mpbp-update}
    \SetAlgoLined
    \DontPrintSemicolon
    
    \For{$k \in \del i$}{
        - Compute $\hat{\mu}_{ki}(\overline{\alpha}_k, \overline{x}_i)$, i.e. the matrix-product state of the coefficients of the basis expansion w.r.t. variables $x_k^t$ (\cref{sec:tt-coefficients}) \;
    }
    \For{$j \in \del i$}{
        - Compute $\hat{\mu}_{\del i \setminus j, i}(\overline{\alpha}_{\del i \setminus j}, \overline{x}_i)$, i.e. the coefficients of the basis expansion of $\mu_{\del i \setminus j, i}(\overline{y}_{\del i \setminus j}, \overline{x}_i) = \int \prod_{k\in \del i \setminus j} \diff{\overline{x}_k}$ $\prod_t p(y_{\del i \setminus j} | \bs{x}_{\del i \setminus j}^t) \prod_{k\in\del i \setminus j} \mu_{ki}(\overline{x}_k, \overline{x}_i)$ w.r.t. variable $y_{\del i \setminus j}$ (\cref{sec:convolution}) \;
        - Compute $\mu_{ij}(\overline{x}_i, \overline{x}_j) = \int \diff{\overline{y}_{\del i \setminus j}}$ $\prod_t p(x_i^{t+1} | y_{\del i \setminus j}^t, x_i^t, x_j^t) \mu_{\del i \setminus j, i}(\overline{y}_{\del i \setminus j}, \overline{x}_i)$ (\cref{sec:hypergeometric}) \;
    }
\end{algorithm}
\end{tcolorbox}

\subsection{Matrix-product states of coefficients}
\label{sec:tt-coefficients}
Take the message $\mu_{ki} (\overline{x}_k, \overline{x}_i) = \prod_t A_{ki}^t (x_k^t, x_i^t)$. The matrix elements $\left[A_{ki}^t (x_k^t, x_i^t)\right]_{m, n}$ (for fixed $x_i^t$) are a discrete function for the spin $x_k^t$ and one wants to calculate the coefficients of its Fourier series. For the sake of generality, a generic function $p(x) = p(-1) \delta_{x,-1} + p(+1) \delta_{x,+1}$ for a spin is considered in the following. For going to the continuum, this is approximated by a sum of two Gaussian functions, centered respectively in $-1$ and $+1$ and with standard deviation $\sigma$:
\begin{equation}
    p(x) \simeq \dfrac{1}{\sqrt{2\pi\sigma^2}} \left( p(-1) e^{-\frac{(x+1)^2}{2\sigma^2}} + p(+1) e^{-\frac{(x-1)^2}{2\sigma^2}} \right).
\end{equation}
Now the coefficients of the Fourier expansion must be calculated. The Fourier series is defined for periodic functions, so one has to choose a period $P$ and redefine the function accordingly (the period must be large enough to safely consider $p(x) = 0$ around $x=-P/2$ and $x=P/2$).\\
Then, the coefficients of the expansion can be calculated as a scalar product with the basis functions:
\begin{equation}
\begin{split}
    \hat{p}(\alpha) =& \dfrac{1}{P} \int_{-P/2}^{+P/2} p(x) u_{\alpha}^*(x)\\
    \simeq& \dfrac{e^{-\frac{1}{2} (\frac{2\pi}{P} \alpha)^2 \sigma^2}}{P} \left[ \left( p(-1) + p(+1) \right) \cos\left( \frac{2\pi}{P} \alpha \right) \right.+\\
    &+ \left. i \left( p(-1) - p(+1) \right) \sin\left( \frac{2\pi}{P} \alpha \right) \right].
\end{split}
\end{equation}
Therefore, one can write the messages in the Fourier basis:
\begin{equation}
    \mu_{ki}(\overline{x}_k, \overline{x}_i) = \prod_t \sum_{\alpha_k^t} \hat{A}_{ki}^t(\alpha_k^t, x_i^t) u_{\alpha_k^t}(x_k^t).
\end{equation}

\subsection{Convolution in the Fourier basis}
\label{sec:convolution}
Having written the messages as a Fourier series, one now needs to compute the messages for the intermediate variables, i.e. $\mu_{A i} (\overline{y}_A, \overline{x}_i)$ for $A \subseteq \partial i$. As explained in \cite{crottiMatrixProductBelief2023c}, this is done recursively, and it is sufficient to define the procedure for computing $\mu_{A\cup B, i} (\overline{y}_{A\cup B}, \overline{x}_i)$ having $\mu_{A i} (\overline{y}_A, \overline{x}_i)$ and $\mu_{B i} (\overline{y}_B, \overline{x}_i)$. In practice, it is necessary to calculate
\begin{equation}
\begin{split}
    \mu_{A\cup B, i} (\overline{y}_{A\cup B}, \overline{x}_i) = \int \diff{\overline{y}_A} \diff{\overline{y}_B} \delta(\overline{y}_{A\cup B} - \overline{y}_A - \overline{y}_B) \\
    \mu_{A i} (\overline{y}_A, \overline{x}_i) \mu_{B i} (\overline{y}_B, \overline{x}_i).
\end{split}
\end{equation}
This can be done straightforwardly; the detailed calculations are shown in \cref{app:convolution}.
The calculation of $\mu_{\{k\}, i} (\overline{y}_{\{k\}}, \overline{x}_i)$ is easy:
\begin{equation}
\begin{split}
    \mu_{\{k\}, i} (\overline{y}_{\{k\}}, \overline{x}_i) =& \int \diff{\overline{x}_k} \delta(\overline{y}_{\{k\}} - J_{ki} \overline{x}_k) \mu_{ki} (\overline{x}_k, \overline{x}_i) \\
    =& \mu_{ki} (\overline{y}_{\{k\}} / J_{ki}, \overline{x}_i).
\end{split}
\end{equation}

\subsection{Calculation of outgoing messages}
\label{sec:hypergeometric}
Using the methodology described above, for each neighbor $j$ of node $i$ one calculates the so-called ``cavity" message $\mu_{\del i \setminus j, i}(\overline{y}_{\del i \setminus j}, \overline{x}_i)$. It remains to calculate the outgoing messages, by including the factors $p(x_i^{t+1} | y_{\del i \setminus j}^t, x_i^t, x_j^t)$:
\begin{equation}
\begin{split}
    \mu_{ij}(\overline{x}_i, \overline{x}_j) = \int \diff{\overline{y}_{\del i \setminus j}} \prod_t & p(x_i^{t+1} | y_{\del i \setminus j}^t, x_i^t, x_j^t) \\
    &\mu_{\del i \setminus j,i}(\overline{y}_{\del i \setminus j}, \overline{x}_i).
\end{split}
\end{equation}
By exploiting the Fourier representation of the cavity messages as described in \cref{sec:convolution}, one has
\begin{align}
    \mu_{ij}(\overline{x}_i, \overline{x}_j) =& \int \diff{\overline{y}_{\del i \setminus j}} \prod_t p(x_i^{t+1} | y_{\del i \setminus j}^t, x_i^t, x_j^t) \nonumber\\
    &\prod_t \sum_{\alpha_{\del i \setminus j}^t} \hat{A}_{\del i \setminus j, i}(\alpha_{\del i \setminus j}^t, x_i^t) u_{\alpha_{\del i \setminus j}^t}(y_{\del i \setminus j}^t) \\
    =& \prod_t \sum_{\alpha_{\del i \setminus j}^t} \hat{A}_{\del i \setminus j, i}(\alpha_{\del i \setminus j}^t, x_i^t) \nonumber\\
    & \int \diff{y_{\del i \setminus j}^t} p(x_i^{t+1} | y_{\del i \setminus j}^t, x_i^t, x_j^t)  u_{\alpha_{\del i \setminus j}^t}(y_{\del i \setminus j}^t) \label{eq:msg-update-int}
\end{align}
and what is left to do is to calculate the integral
\begin{equation}
    I_{\gamma} (x_i^{t+1}, x_j^t) \coloneqq \int_{-P/2}^{+P/2} \diff{y} p(x_i^{t+1} | y, x_i^t, x_j^t)  u_{\gamma}(y) \label{eq:integral-hypergeometric}
\end{equation}
which has a closed-form expression involving the Gausssian hypergeometric function, as shown in detail in \cref{app:integral-hypergeometric}.
By plugging it into \cref{eq:msg-update-int}, one obtains:
\begin{equation}
\begin{split}
    \mu_{ij}(\overline{x}_i, \overline{x}_j) =& \prod_t \sum_{\alpha_{\del i \setminus j}^t} \hat{A}_{\del i \setminus j, i}(\alpha_{\del i \setminus j}^t, x_i^t) I_{\alpha_{\del i \setminus j}} (x_i^{t+1}, x_j^t) \\
    =& \prod_t B_{ij}(x_i^{t+1}, x_i^t, x_j^t).
\end{split}
\end{equation}
The final sought form is $\mu_{ij}(\overline{x}_i, \overline{x}_j) = \prod_t A_{ij}(x_i^t, x_j^t)$. In the previous form, the variables $x_i^{t+1}$ and $x_i^t$ were coupled in the same tensor $B_{ij}^t$. To solve this, a sweep of SVD is sufficient.
\\
The computational cost per iteration for Kinetic Ising scales as $\mathcal{O}((d^6 K + d^2K^2)|E|T)$, where $|E|$ is the number of edges in the graph, $d$ is the bond dimension of the matrices, $K$ is the dimension of the basis, and $T$ is the time horizon.

\section{Results}
\label{sec:results}
We validate Basis-MPBP on the  Kinetic Ising model by comparing its predictions with exact solutions or results of Monte Carlo simulations, when any of the two are possible to retrieve. As already mentioned, the standard implementation of MPBP becomes impractical for nodes of (moderately) high degree with continuous, disordered couplings with their neighbors. In the following numerical tests, we take the values of the couplings $J_{ki}$ to be independent random variables uniformly distributed in some interval. All code used to produce the results is accessible at \cite{florioFedericoFlorioMatrixProductBPjl2025}. For convenience, we report in \cref{app:parameters} the parameters used in all experiments.

\subsection{Comparison with exact solution}
\label{sec:results-exact}
\begin{figure}[t]
    \centering
    \includegraphics[width=\columnwidth]{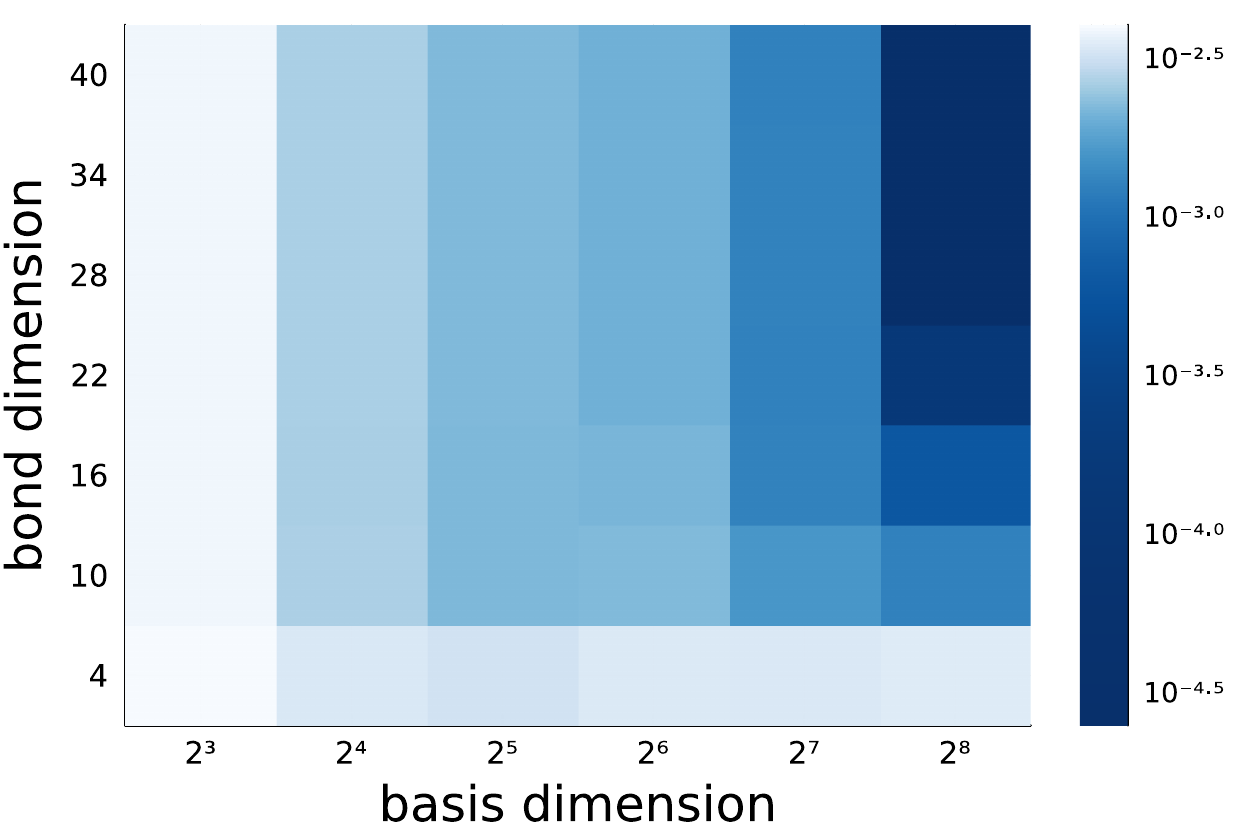}
    \caption{Comparison between Basis-MPBP and exact solution on a tree of $N=10$ nodes. $\beta=0.5$, $J_{k,i} = J_{i,k} \sim \text{Uniform}[-1,+1]$, $h_i = 0$, $m^0 = 0.7$. The algorithm was run with several values of $d$ and $K$. The plot shows the base-10 logarithm of the error $\text{Er}(d,K) = \sqrt{\frac{1}{NT} \sum_{i,t} \left( m_i^t - \tilde{m}_i^t \right)^2}$}
    \label{fig:res-convergence}
\end{figure}
In very small systems, an exact solution can be calculated by computing the probability of every single state of the system and evolving it with the Markov transition matrix. This allows us to evaluate, in a simple case, the errors introduced by all the approximations of Basis-MPBP. In order to evaluate the errors exclusively due to the matrix-product approximation and the basis expansion, we consider a tree graph, for which the cavity approximation is exact.
Considering a tree graph of $N=10$ nodes with symmetric random couplings between the nodes distributed uniformly in $[-1,+1]$, we calculate exactly the magnetizations $m_i^t = \sum_{s=\pm 1} s \prob{x_i^t=s}$ of all nodes at all times $t\in\{0,\dots,10\}$, for a free Kinetic Ising dynamics with time horizon $T=4$ and compare them with the same quantities calculated by means of Basis-MPBP $\tilde{m}_i^t$. The inverse temperature is $\beta=0.5$, there is no external field $h=0$ and the initial condition is given by $f_i^0(x_i^0) = \frac{1}{2}(1+x_i^0 m^0)$ (so that the initial magnetization is $m^0=0.7$). In \cref{fig:res-convergence} we show the root-mean-square error $\text{Er}(d,K) = \sqrt{ \frac{1}{NT} \sum_{i,t} \left( m_i^t - \tilde{m}_i^t \right)^2}$ for different values of the bond dimension $d$ and the basis size $K$. As can be seen, the error decreases when the parameters are increased.

\subsection{Comparison with Monte Carlo}
\label{sec:results-mc}

\begin{figure}[t]
    \centering
    \begin{subfigure}{0.4\textwidth}
        \centering
        \includegraphics[width=\textwidth]{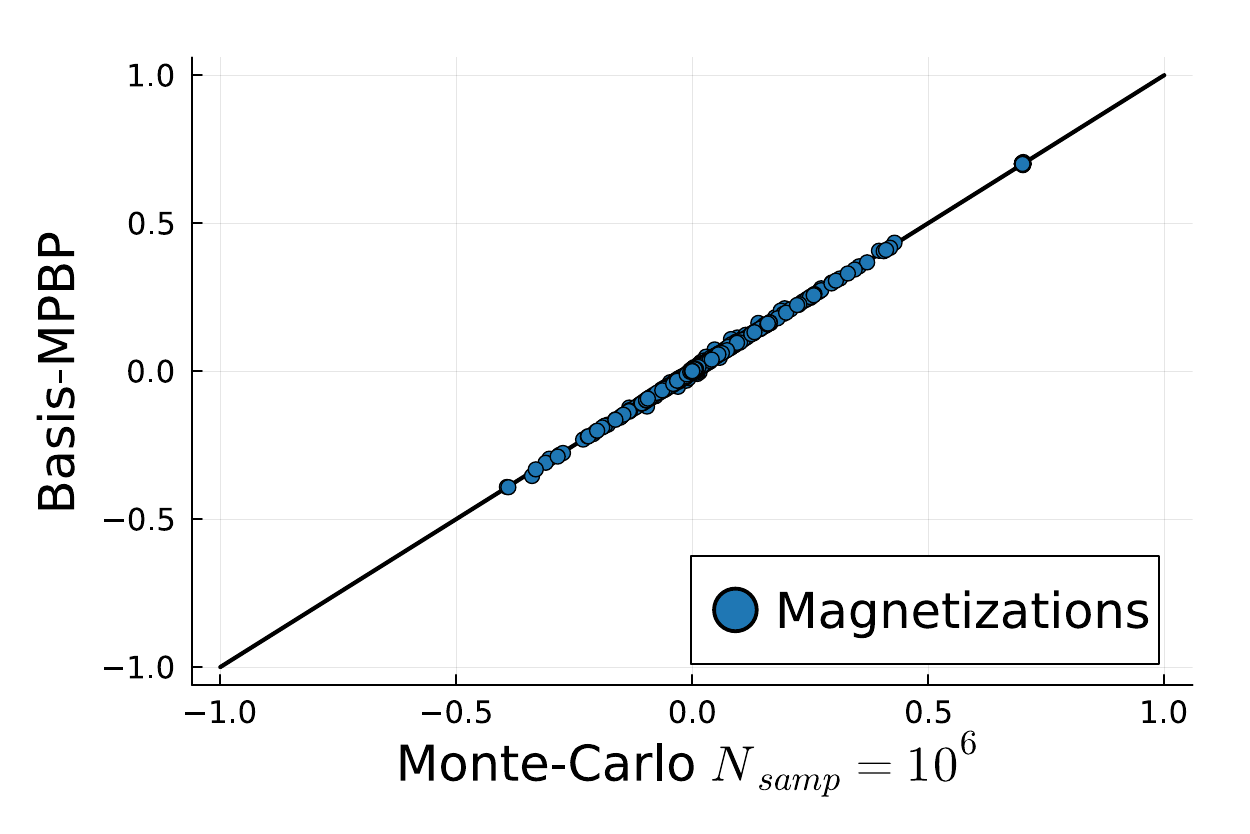}
        \label{fig:res-tree-energy}
    \end{subfigure}
    \hfill
    \begin{subfigure}{0.4\textwidth}
        \centering
        \includegraphics[width=\textwidth]{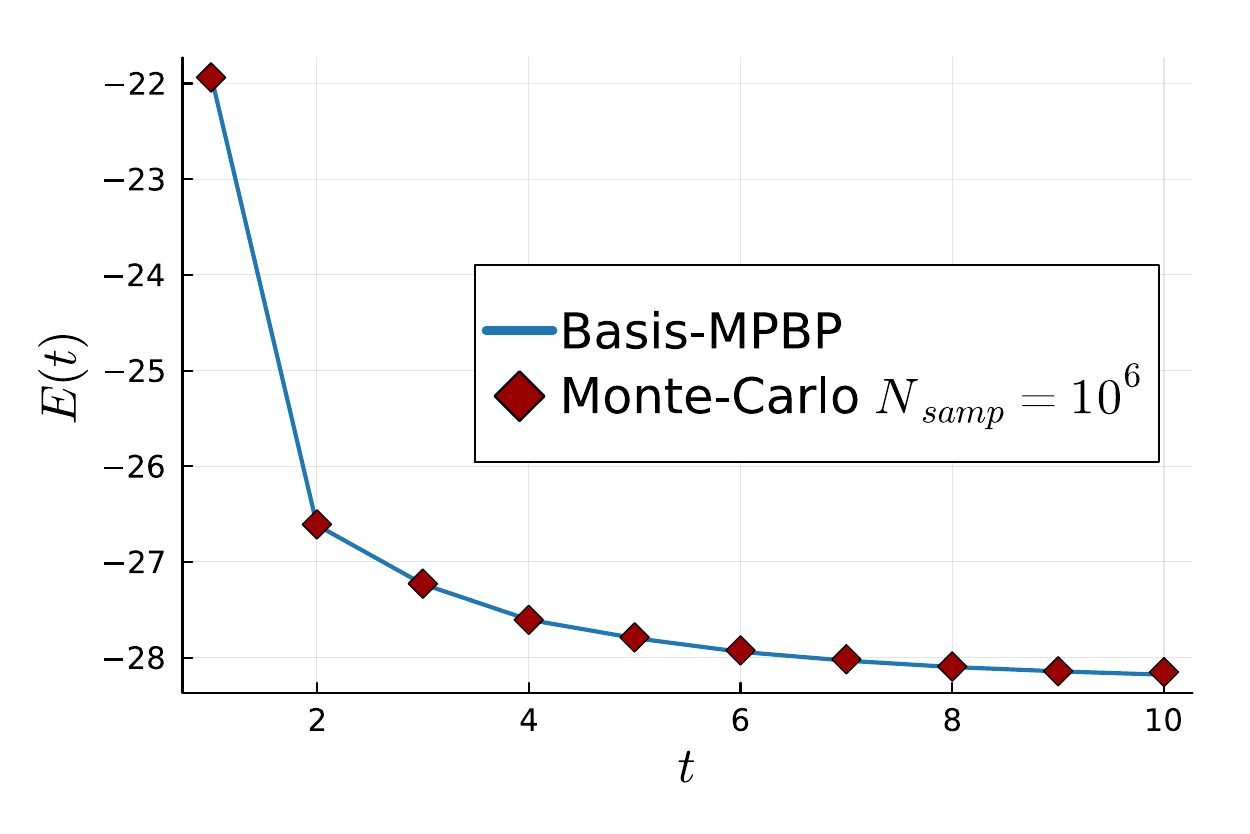}
        \label{fig:res-tree-scatter}
    \end{subfigure}
    
    \caption{Random tree ($N=100$) built with preferential attachment. $\beta=0.5$, $J_{k,i} = J_{i,k} \sim \text{Uniform}[-1,+1]$, $h_i = 0$, $m^0 = 0.7$. The algorithm was run with $d=15$ and $K=100$. Top: magnetizations of all nodes at all times (Monte Carlo vs MPBP with Fourier decomposition).
    Bottom: Time evolution of the ``time-shifted energy'' of the system.}
    \label{fig:res-tree}
\end{figure}

\begin{figure}[t]
    \centering
    \begin{subfigure}{0.4\textwidth}
        \centering
        \includegraphics[width=\textwidth]{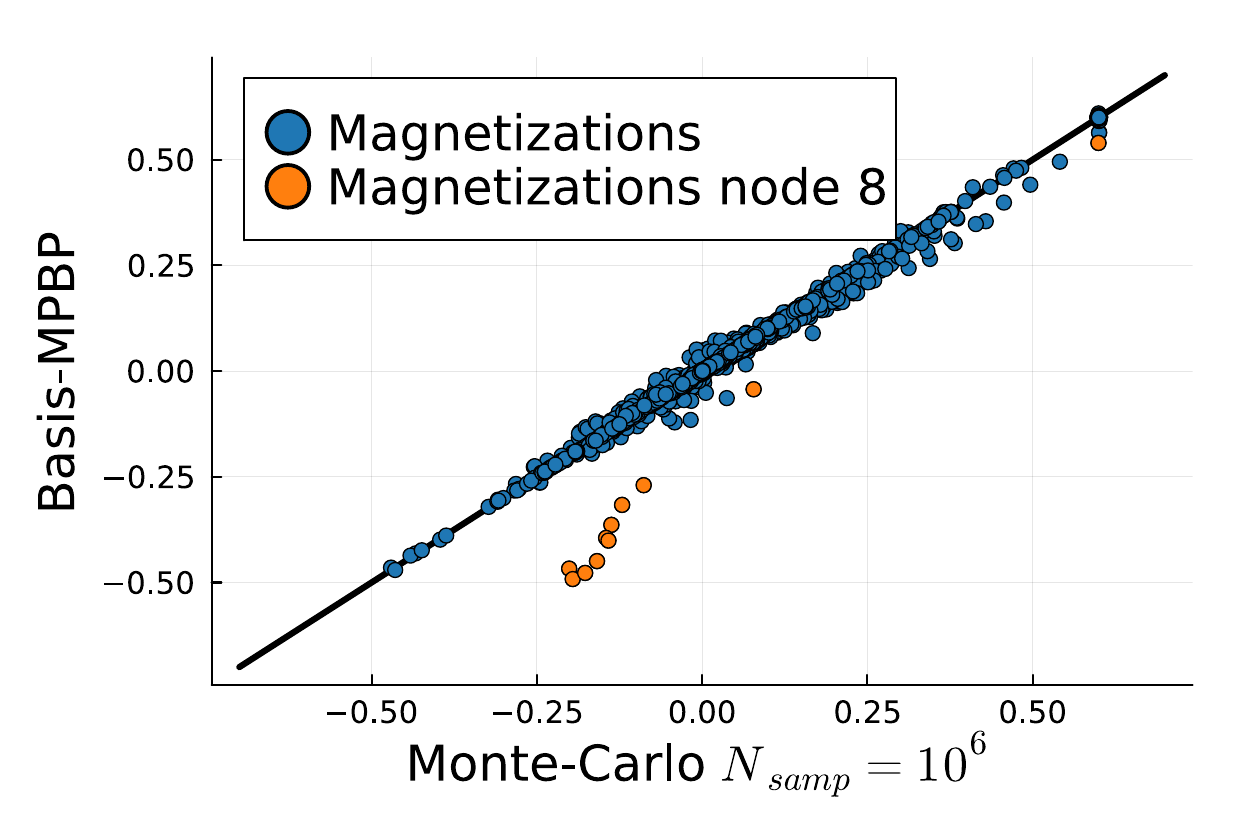}
        \label{fig:res-conf-energy}
    \end{subfigure}
    \hfill
    \begin{subfigure}{0.4\textwidth}
        \centering
        \includegraphics[width=\textwidth]{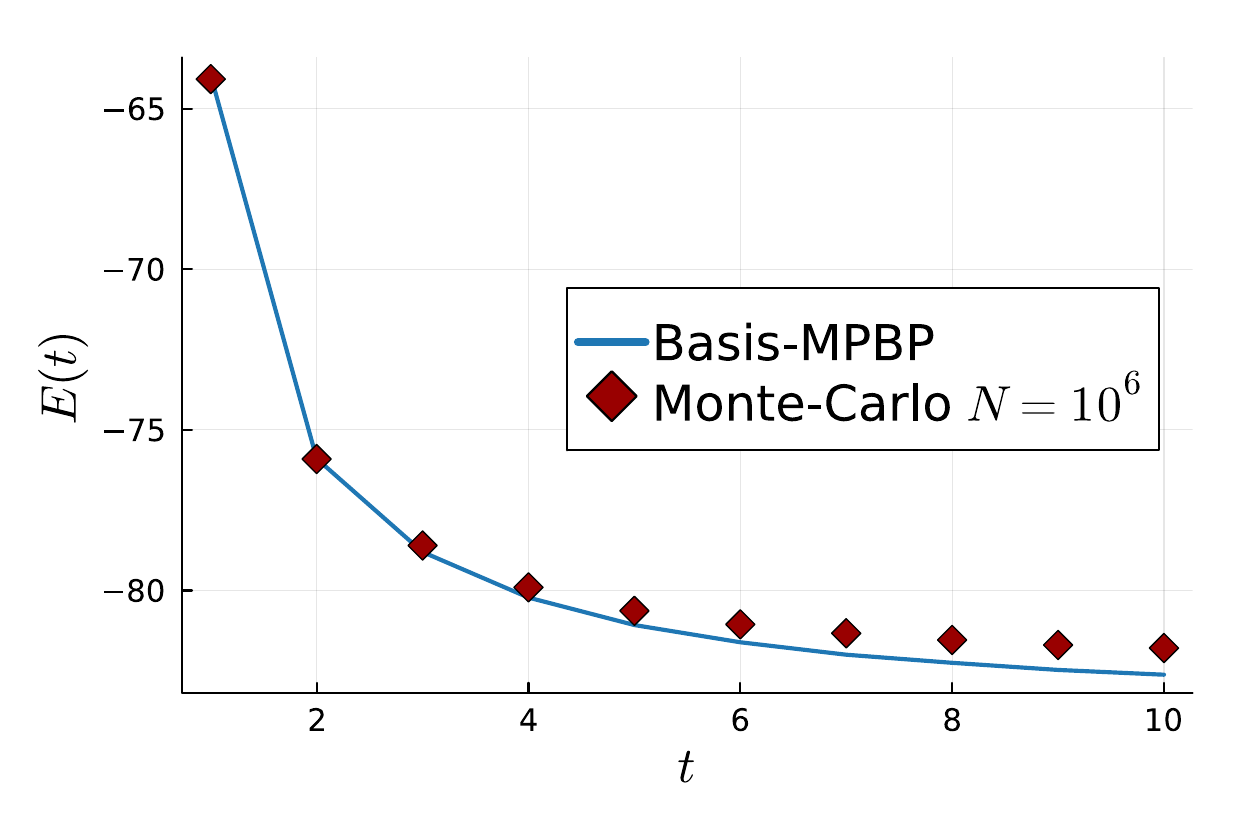}
        \label{fig:res-conf-scatter}
    \end{subfigure}
    
    \caption{Random graph ($N=200$) built with configuration model. $\beta=0.7$, $J_{k,i} = J_{i,k} \sim \text{Uniform}[-1,+1]$, $h_i = 0$, $m^0 = 0.6$. The algorithm was run with $d=20$ and $K=100$. Top: magnetizations of all nodes at all times (Monte Carlo vs MPBP with Fourier decomposition).
    Bottom: Time evolution of the ``time-shifted energy'' of the system.}
    \label{fig:res-conf}
\end{figure}

\begin{figure*}[ht]
    \centering
    \begin{subfigure}{0.49\textwidth}
        \centering
        \includegraphics[width=\textwidth]{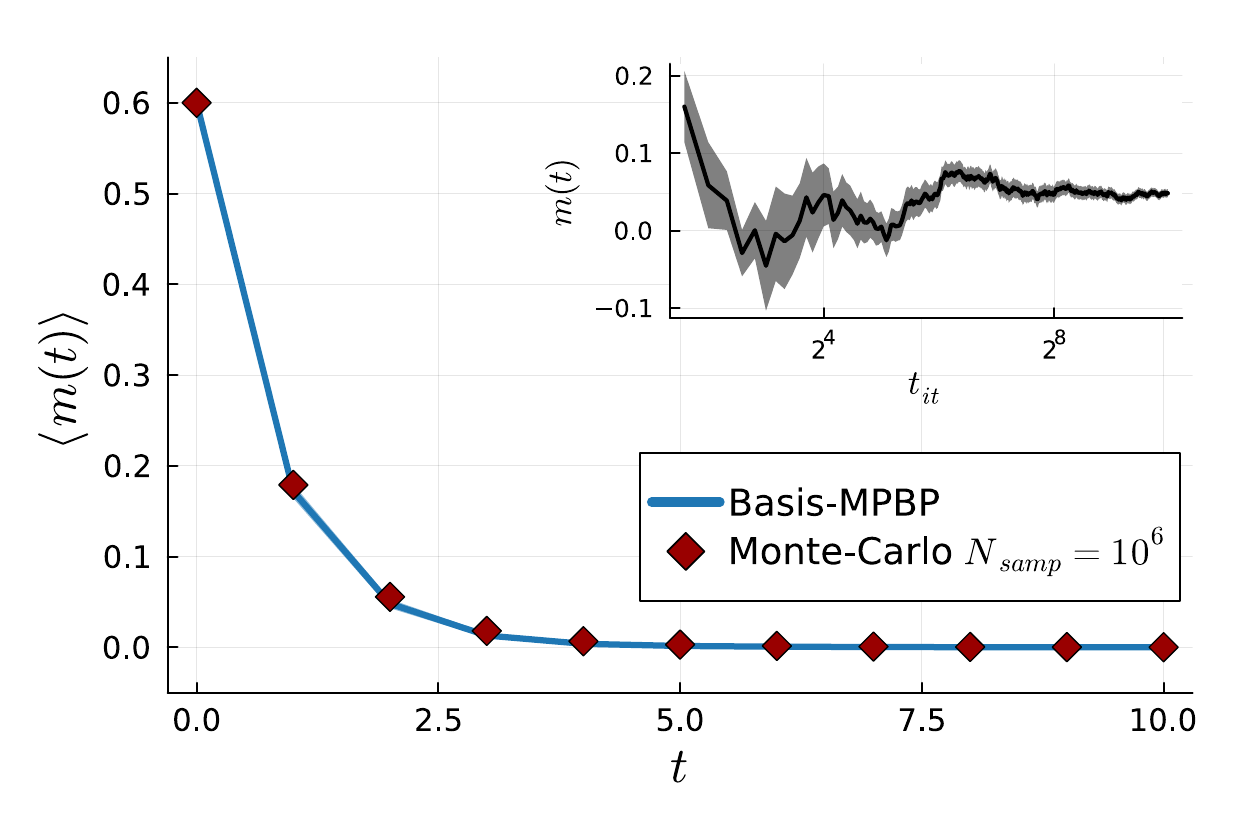}
        \label{fig:res-rr-magnetization}
    \end{subfigure}
    \hfill
    \begin{subfigure}{0.49\textwidth}
        \centering
        \includegraphics[width=\textwidth]{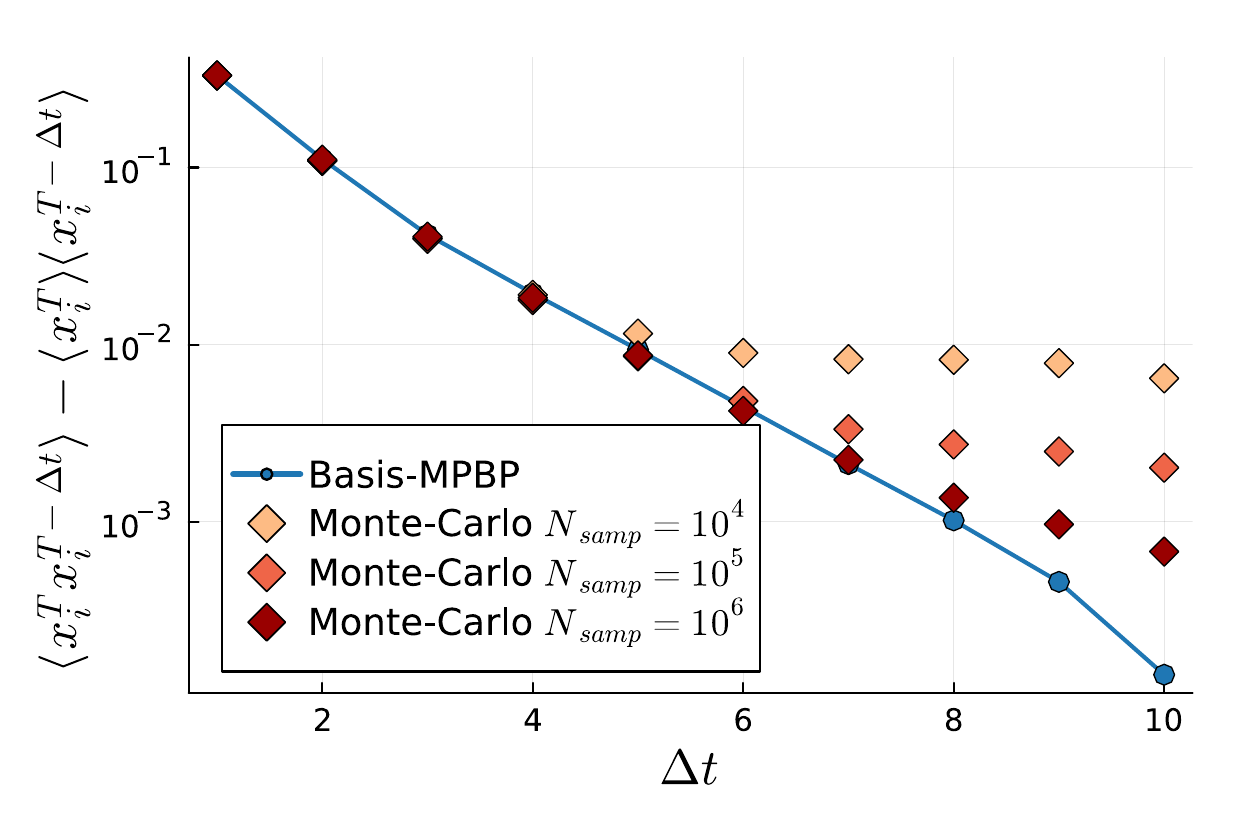}
        \label{fig:res-rr-autocorrelation}
    \end{subfigure}    
    \caption{Average magnetization (left) and auto-correlation (right) on an infinite random-regular graph with degree 8 and with $\beta=0.3$, $J_{k,i} \sim \text{Uniform}[-1,+1]$, $h=0$, $m^0=0.6$. The algorithm was run with $d=15$ and $K=60$. The inset shows the evolution of the average magnetization at time $t=3$ as a function of the simulation time $t_{\text{it}}$.}
    \label{fig:res-rr}
\end{figure*}

\begin{figure*}[ht]
    \centering
    \begin{subfigure}{0.49\textwidth}
        \centering
        \includegraphics[width=\textwidth]{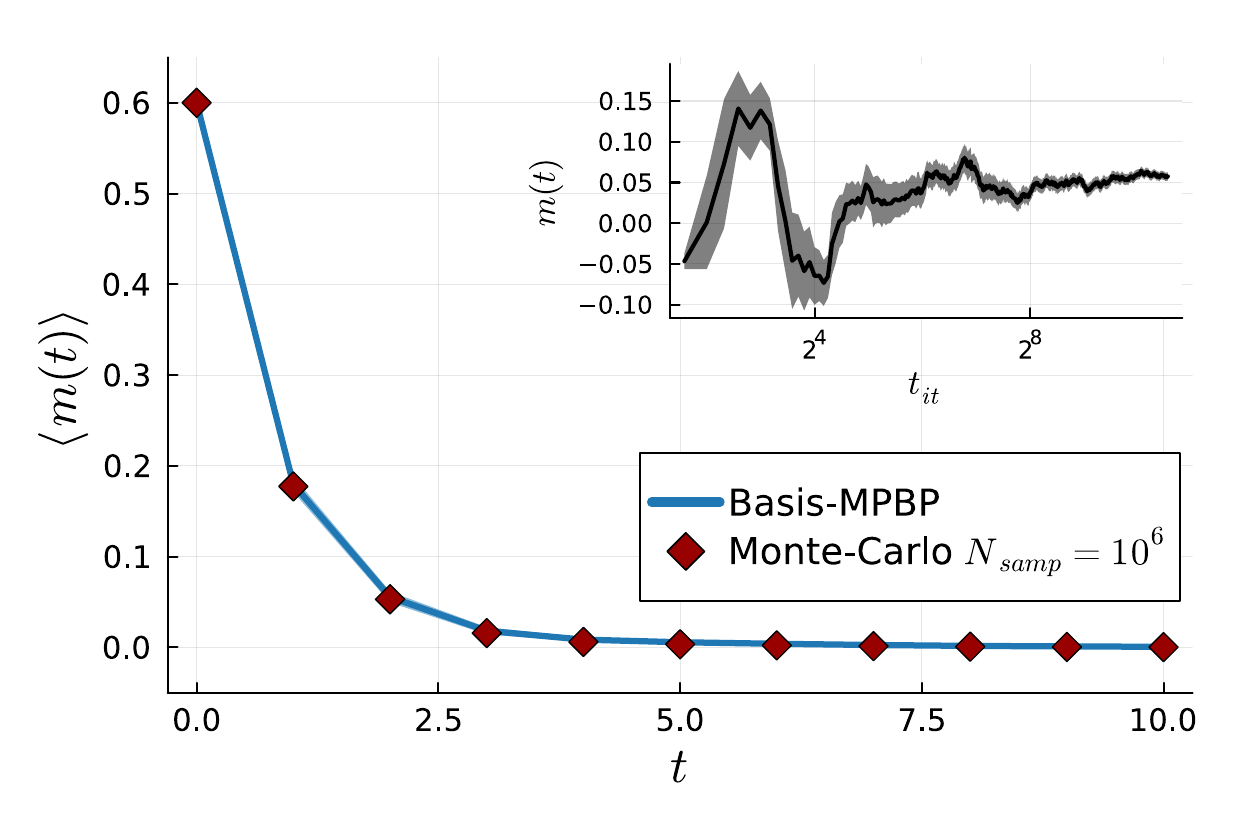}
        \label{fig:res-er-magnetization}
    \end{subfigure}
    \hfill
    \begin{subfigure}{0.49\textwidth}
        \centering
        \includegraphics[width=\textwidth]{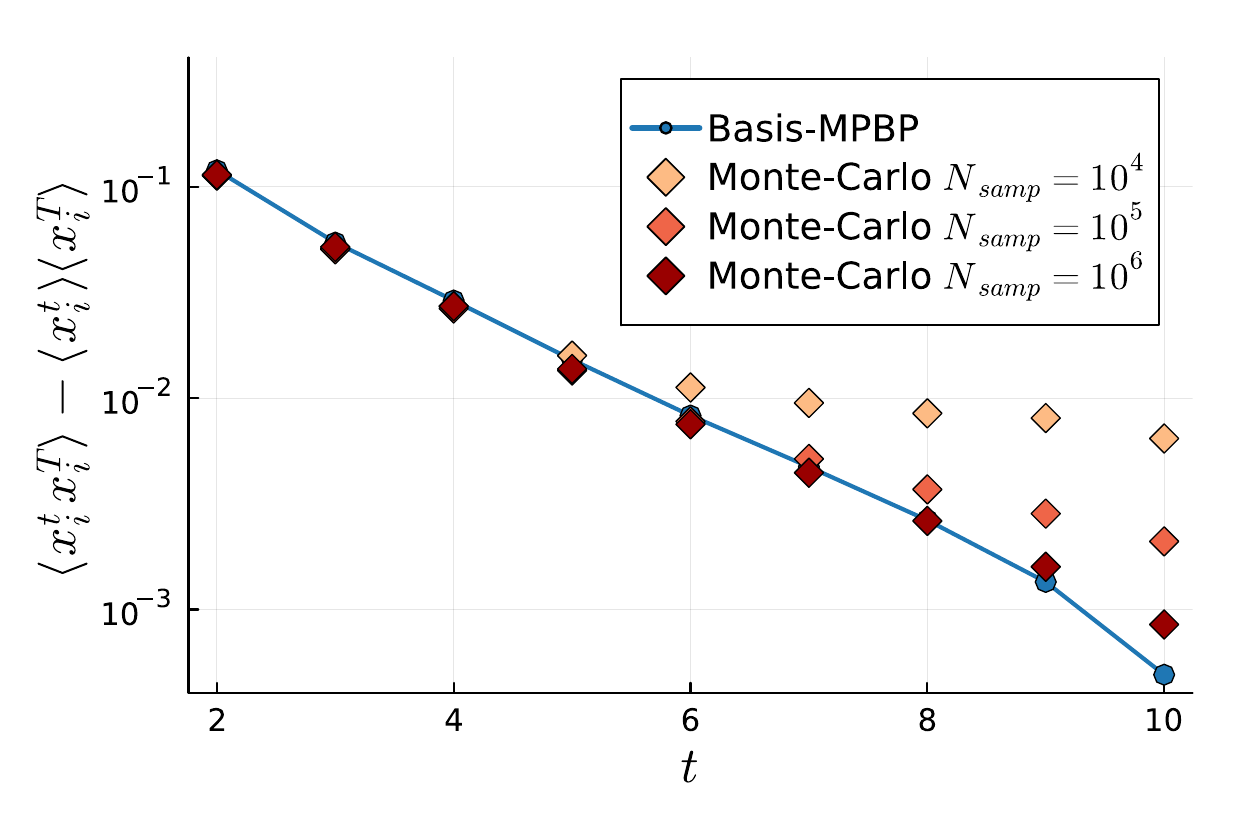}
        \label{fig:res-er-autocorrelation}
    \end{subfigure}   
    \caption{Average magnetization (left) and auto-correlation (right) on an infinite Erdős-Rényi graph with average degree 4 and with $\beta=0.5$, $J_{k,i} \sim \text{Uniform}[-1,+1]$, $h=0$, $m^0=0.6$. The algorithm was run with $d=15$ and $K=60$. The inset shows the evolution of the average magnetization at time $t=3$ as a function of the simulation time $t_{\text{it}}$.}
    \label{fig:res-er}
\end{figure*}

Next, we compare the results of Basis-MPBP with Monte Carlo simulations. This makes it possible to treat larger systems, for which the exact solution would be impossible to compute.
For a first validation, we employ a tree graph of size $N=100$, built with preferential attachment so as to include high-degree nodes. Again, on a tree, the DBP equations are exact, so we can estimate the error in the approximation exclusively due to the matrix-product \textit{ansatz} and the basis expansion.
The inverse temperature is $\beta=0.5$, the couplings are symmetric $J_{k,i} = J_{i,k} \sim \text{Uniform}[-1,+1]$, no external field is present and $m^0=0.7$. The marginals of all nodes at all times are computed, both with Basis-MPBP and with Monte Carlo simulations. We use parameters $d=15$ and $K=100$, which, taking \cref{fig:res-convergence} as a rule of thumb, should give an error of approximately $10^{-3}$, comparable to the sampling error of $10^6$ Monte Carlo samples. The results are shown in \cref{fig:res-tree}, together with the time evolution of the ``time-shifted energy'' of the system $E(t) = -\sum_{i,j}  J_{ji} \left\langle x_i^{t+1} x_j^t \right\rangle$. Since the couplings are symmetric, $J_{k,i}=J_{i,k} \forall i,k$, in the infinite-time limit this quantity converges to the equilibrium energy $-\sum_{i,j} J_{i,j} \langle x_i x_j \rangle_{\text{eq}}$. This holds because the equilibrium joint distribution for two neighboring variables $p^{\text{eq}}(x_i,x_j)$ is equal to $p(x_i^{t+1}, x_j^t)$, where the configurations $\bs{x}^{t+1}$ and $\bs{x}^t$ are sampled using the Kinetic Ising dynamics at the steady state (see the appendix of \cite{crottiMatrixProductBelief2023c} for a proof).\\
The agreement between the two methods is good. Indeed, the average error on the magnetizations calculated with Monte Carlo $m_i^t$ and Basis MPBP $\tilde m_i^t$ is $\sqrt{\frac{1}{NT} \sum_{i,t} \left( m_i^t - \tilde{m}_i^t \right)^2} = 3.3\times 10^{-3}$, which is compatible with the Monte Carlo sampling error.\\
Next, the same test is performed on a graph of $N=200$ nodes with a power-law degree distribution $p(c) \sim c^{-0.6}$, built with a configuration model. For this simulation we use parameters $d=20$ and $K=100$. In this graph there are loops, including several short ones. Nevertheless, as can be seen in \cref{fig:res-conf}, the results show again a good agreement with Monte Carlo, with an exceptional deviation for a specific node (node 8, which is the eighth node in order of degree and participates in a large number of very short loops). In this extremely loopy graph, the root-mean-square error with respect to Monte Carlo is $2.0\times10^{-2}$ and the energy of the system is slightly underestimated by Basis-MPBP.

\subsection{Infinite systems}
\label{sec:results-infinite}

\begin{figure*}[htbp]
    \centering
    \begin{subfigure}{0.51\textwidth}
        \centering
        \includegraphics[width=\textwidth]{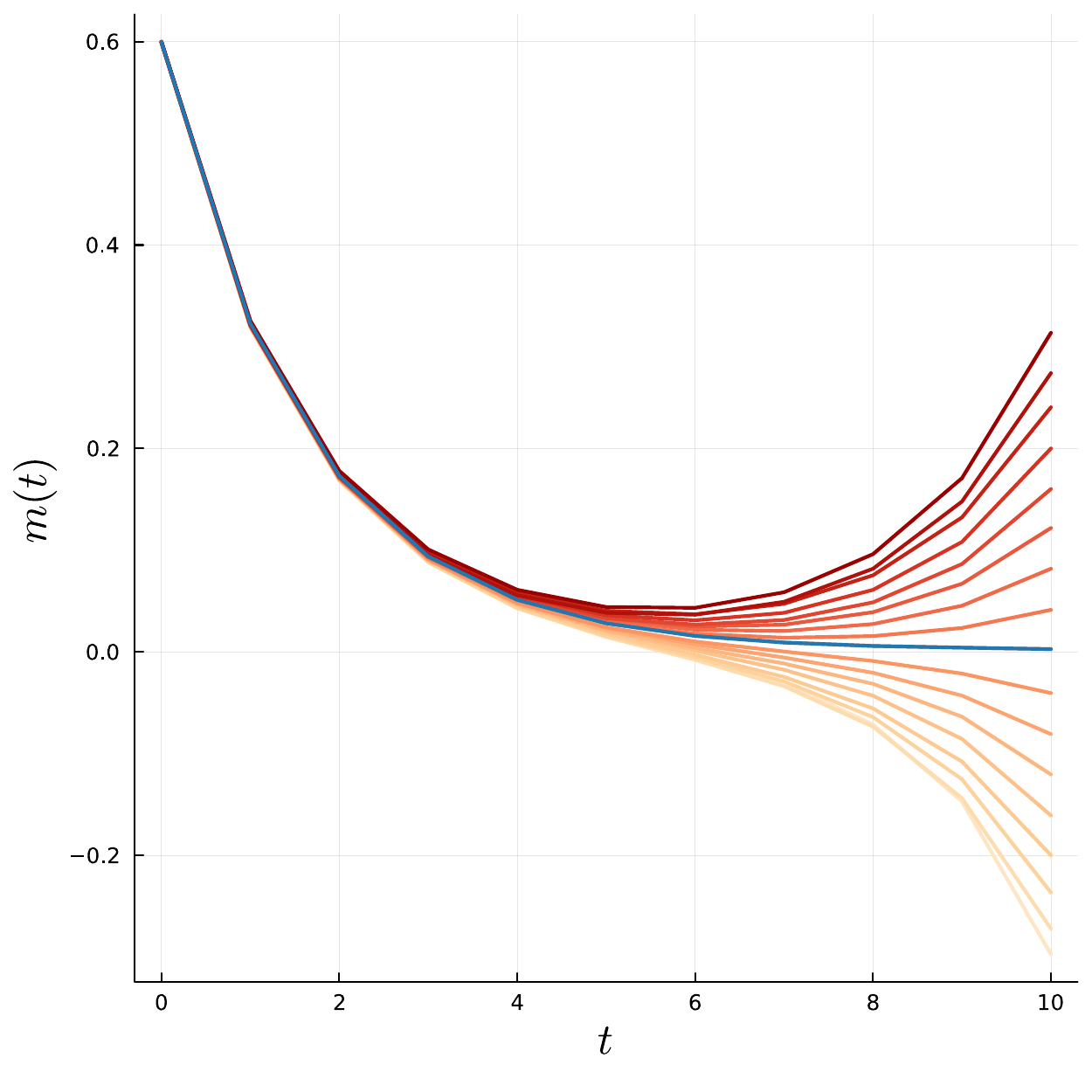}
        \label{fig:res-ld-magnetizations}
    \end{subfigure}
    \hfill
    \begin{subfigure}{0.48\textwidth}
    \begin{subfigure}{\textwidth}
        \centering
        \includegraphics[width=\textwidth]{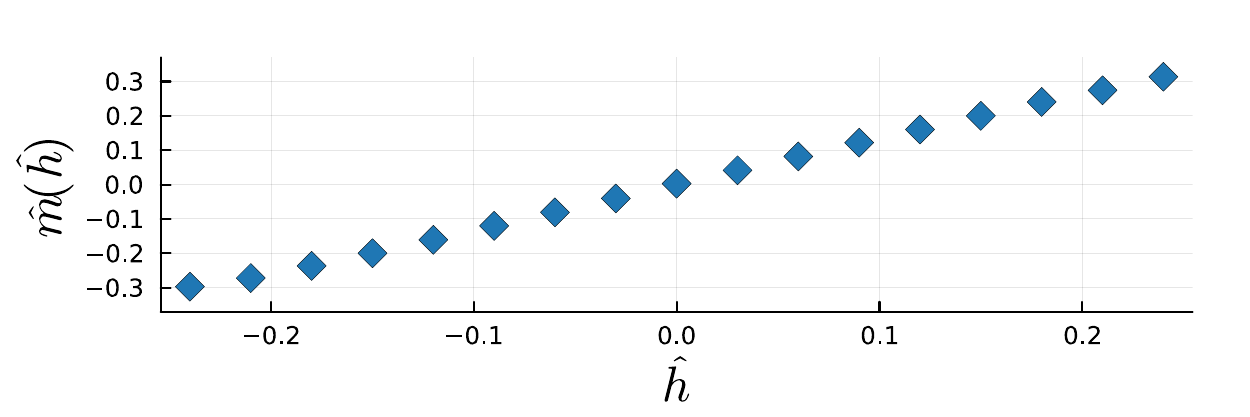}
        \label{fig:res-ld-m(h)}
    \end{subfigure}
    \begin{subfigure}{\textwidth}
        \centering
        \includegraphics[width=\textwidth]{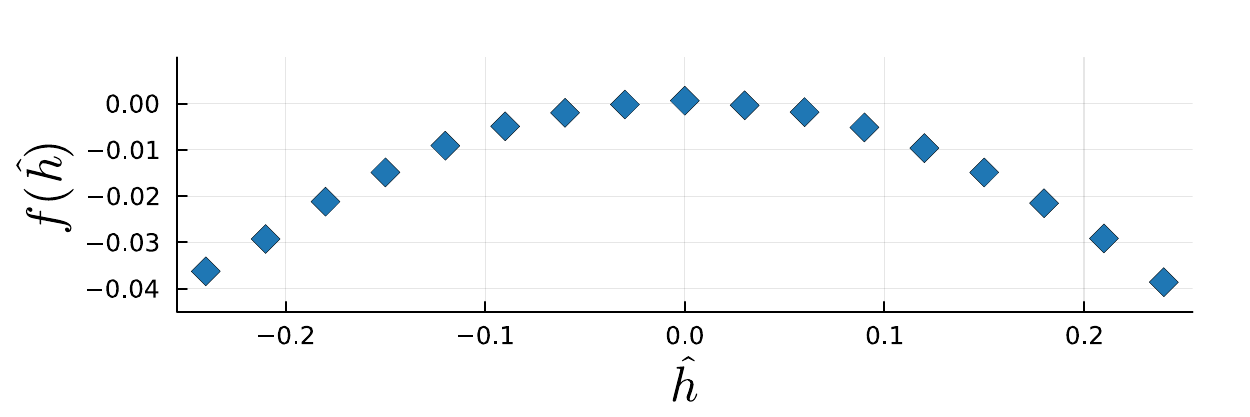}
        \label{fig:res-ld-f(h)}
    \end{subfigure}
    \begin{subfigure}{\textwidth}
        \centering
        \includegraphics[width=\textwidth]{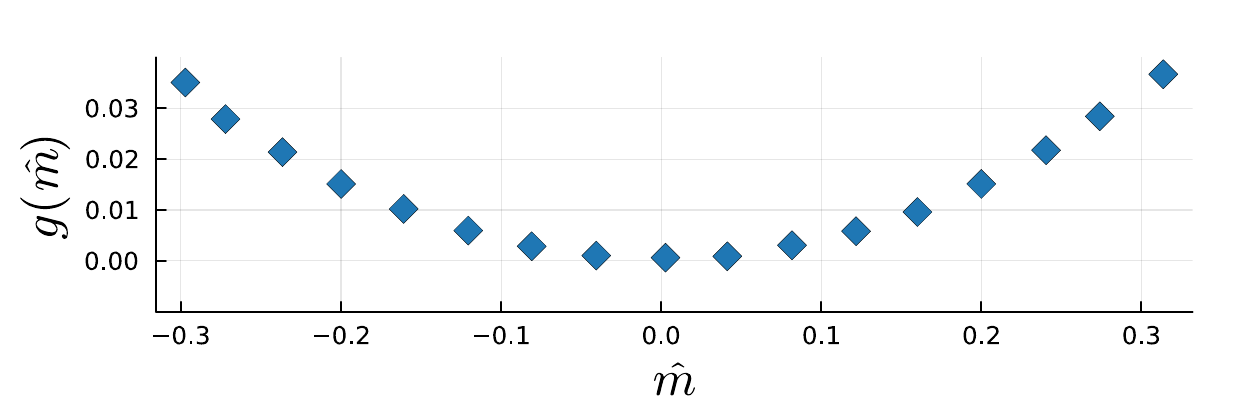}
        \label{fig:res-ld-g(m)}
    \end{subfigure}
    \end{subfigure}
    
    \caption{Trajectories in time (left), final-time magnetizations (right top), free energies (right middle) and large deviations law (right bottom) on an infinite regular graph with degree 8 and with $\beta=1/7$, $J_{k,i} \sim \text{Uniform}(0, 1)$, $h=0$, $m^0=0.6$. The values of the reweighting field $\hat{h}$ are between $-0.3$ and $0.3$. The algorithm was run with $d=10$ and $K=30$.}
    \label{fig:res-ld}
\end{figure*}

Then, we consider the Kinetic Ising dynamics in a case without equilibrium (asymmetric couplings) on infinite graphs. \Cref{fig:res-rr,fig:res-er} show the average magnetization, the auto-correlation and the average energy of an infinite random regular graph with degree $8$ and an infinite Erdős-Rényi graph with average degree 4, respectively. The results are obtained with population dynamics (see \cref{app:popdyn}), using populations of 100 messages initialized as matrix-product states with matrix entries uniformly drawn at random in the interval $[0,1]$. Here, the inertial parameter in \cref{eq:kineticising} is set to $\rho=0.3$.\\
For each iteration $t_{\text{it}}$, we compute the average and standard deviation of some selected observables over the last $t_{\text{it}}/2$ iterations and we stop the algorithm when the observables are visually stable. As an example, in the inset of \cref{fig:res-rr,fig:res-er} we show the evolution of the average magnetization of the system at time $t=2$.\\
The results show that estimating small auto-correlations at large $\Delta t$ requires a very large number of Monte Carlo samples. The reason is that an auto-correlation of order $10^{-4}$ is obtained by averaging random variables of order unity (here, products taking values $\pm1$). Consequently, resolving a mean of order $10^{-4}$ with a relative error of 10\%, corresponding to an absolute error of order $10^{-5}$, requires approximately $10^{10}$ independent samples.

Finally, a study is performed regarding large deviations from the typical trajectories. The aim is to compute the large deviation law for $\hat{m} = \frac{1}{N} \sum_i x_i^T$, the magnetization at time $T$. In particular, we seek $g(\hat{m}) \coloneqq -\frac{1}{N} \log  \sum_{\overline{\bs{x}}} p(\overline{\bs{x}}) \delta_{\hat{m}, \frac1N \sum_i x_i^T}$ so that the probability to observe a magnetization $\hat{m}$ at time $T$ is $p(\hat{m}) = e^{-N g(\hat{m})}$. Further details and a validation study on a small graph are in \cref{app:large-deviations}.
\Cref{fig:res-ld} shows, on an infinite random-regular graph with degree 8, with $\beta = 1/7$ and couplings drawn uniformly at random from $[0,1]$ (which is in the paramagnetic phase at equilibrium), the time trajectories of the magnetization, the final-time magnetization $\hat m(\hat h)$ and Bethe free energies $f(\hat h)$ as a function of the reweighting field, and $g(\hat{m})$, obtained as a parametric plot $g(\hat{m}(\hat{h})) = f(\hat{h})+\hat{h}\hat{m}(\hat{h})$ vs. $\hat{m}(\hat{h})$. These results have been obtained by the population dynamics implementation of Basis-MPBP described in \cref{app:popdyn}, using a population of 100 messages initialized as matrix-product states with matrix entries uniformly drawn at random in the interval $[0,1]$. Let us emphasize that this study would be virtually impossible to perform with Monte Carlo, even for moderately large $N$, as it would require a number of samples that is exponentially large in $N$. As an example, the probability of obtaining one sample with magnetization around $\hat{m}=0.3$ is $p(\hat{m})=e^{-N g(\hat{m})}$ which can be estimated from \cref{fig:res-ld}; for $N=10^3$, one would require around $e^{N g(\hat{m})}\approx 10^{13}$ samples to obtain a single sample with the target $\hat{m}$, or around $10^{17}$ samples to achieve an error of $0.01$ for that $\hat{m}$. Note that in this disordered model, the estimation with plain MPBP (as opposed to Basis-MPBP) would also be prohibitive for nodes of degree 8.

\section{Conclusion}
This work proposes Basis-MPBP, an extension of MPBP \cite{crottiMatrixProductBelief2023c}, based on the matrix-product \text{ansatz}, to implement the Dynamic Belief Propagation scheme on continuous variables. Like the original implementation of MPBP, it can be applied both to free and reweighted dynamics, allowing to treat both conditioned dynamics and rare events. While in principle the MPBP iteration for one node on general models is exponential in the node's degree, in \cite{crottiMatrixProductBelief2023c} a family of models is presented for which this iteration can be made linear in the node's degree (and thus a full iteration is linear in the number of edges). In this work, we present an alternative formulation of MPBP, based on an alternative factor graph, which helps to clarify which models this efficient iteration can be applied to. This new formulation also helps to simplify the presentation of Basis-MPBP.

The accuracy of Basis-MPBP is controlled by two parameters, namely the bond dimension $d$ and the basis size $K$. While it becomes exact on acyclic graphs only when $d,K\to \infty$, small values of  $d$ and $K$  are typically sufficient to obtain good accuracy.  For fixed values of the parameters, the iteration is linear both in the number of edges $|E|$ and in the time horizon $T$, albeit with a relatively large constant. Basis-MPBP can also be applied to study random graphs in the infinite size limit in a cavity approach, where a distribution of messages is parametrized by a population of matrix-product states.\\
Basis-MPBP can be applied to fully continuous-state-space models and mixed continuous/discrete ones, in which expansion in the basis is done only to account for the dependence on the continuous variables. We applied it to the Kinetic Ising model with real-valued disordered couplings, for which the local fields are effectively continuous. We validated it against an exact solution and Monte Carlo simulations, with excellent agreement on quantities for which MCMC estimation is feasible. On some observables, such as auto-correlations, Monte Carlo estimates are subject to large stochastic error and require a prohibitively large sample size. This problem is also present when evaluating conditioned or reweighted dynamics such as those needed to study large deviations. Monte Carlo estimation is extremely inefficient there, as the relevant trajectories in the reweighted dynamics have exponentially small probability to be observed in the free one. 
Basis-MPBP can be easily applied to study conditioned dynamics (as e.g. in \cite{braunsteinSmallcouplingDynamicCavity2025}) and an extension to the analysis of the steady-state can be envisaged, in which messages are infinite matrix-product states parametrized by a unique matrix (see \cite{crottiNonequilibriumSteadystateDynamics2025a}).

\section*{Acknowledgments}
This study was carried out within the FAIR - Future Artificial Intelligence Research project and received funding from the European Union Next-GenerationEU
(Piano Nazionale di Ripresa e Resilienza (PNRR)–Missione 4 Componente 2, Investimento 1.3–
D.D. 1555 11/10/2022, PE00000013). This manuscript reflects only the authors’ views and
opinions; neither the European Union nor the European Commission can be considered responsible for them. We
acknowledge support from the European REA, Marie Skłodowska-Curie Actions, grant agreement no. 101131463 (SIMBAD).

\section*{Data Availability}
Open-source code for the method and all numerical simulations used to produce the results in the manuscript is publicly available at \cite{florioFedericoFlorioMatrixProductBPjl2025}.

\clearpage

\begin{onecolumngrid}
\appendix

\section{Detailed calculation for MPBP update step under basis expansion}\label{app:mpbp-basis}
The detailed calculation of the BP update equation under the basis-expanded-matrix-product \textit{ansatz} $\mu_{ki} \left( \overline{x}_{k}, \overline{x}_{i} \right) = \prod_{t=0}^{T} \sum_{\alpha_k^t, \alpha_i^t} A_{ki}^{t} [\alpha_k^t, \alpha_i^t] u_{\alpha_k^t}\left(x_k^t\right) u_{\alpha_i^t}\left(x_i^t\right)$ is shown here. Only the case of two incoming messages is presented, as the calculation for more than two is done recursively, as introduced in \cite{crottiMatrixProductBelief2023c}.\\
The calculation below considers the case in which all the dynamical variables are continuous. However, if the model includes some discrete variables, $x_i^t\in\{q_1, \dots, q_m\}$, the same procedure can be employed formally by considering bases of Dirac delta functions $u_\alpha(x) = \delta(x-q_\alpha)$ for them.
\begin{align*}
    \mu_{ij} \left( \overline{x}_{i}, \overline{x}_{j} \right) =& \prod_{t=0}^{T} \int_{[a,b]^{2}} \text{d} x_{1}^{t} \text{d} x_{2}^{t} f_{i}^{0} \left( x_{i}^{0} \right) \prod_{t=1}^{T} f_{i}^{t+1} \left(x_{i}^{t+1}, x_{i}^{t}, x_{j}^{t}, x_{1}^{t}, x_{2}^{t} \right) \mu_{1 i} \left( \overline{x}_{1}, \overline{x}_{i} \right) \mu_{2 i} \left( \overline{x}_{2}, \overline{x}_{i} \right) \\
    =& \prod_{t=0}^{T} \int_{[a,b]^{2}} \text{d}x_{1}^{t} \text{d}x_{2}^{t} f_{i}^{0} \left( x_{i}^{0} \right) \prod_{t=0}^{T-1} f_{i}^{t+1} \left(x_{i}^{t+1}, x_{i}^{t}, x_{j}^{t}, x_{1}^{t}, x_{2}^{t} \right) \prod_{t=0}^{T} A_{1 i}^{t} \left( x_{1}^{t}, x_{i}^{t} \right) \prod_{t=0}^{T} A_{2 i}^{t} \left( x_{2}^{t}, x_{i}^{t} \right)\\
    =& \prod_{t=0}^{T} \left(\int_{[a,b]^{2}} \text{d}x_{1}^{t} \text{d}x_{2}^{t} f_{i}^{t+1} \left( x_{i}^{t+1}, x_{i}^{t}, x_{j}^{t}, x_{1}^{t}, x_{2}^{t} \right) A_{1 i}^{t} \left( x_{1}^{t}, x_{i}^{t} \right) \otimes A_{2 i}^{t} \left( x_{2}^{t}, x_{i}^{t} \right) \right)\\
    =& \prod_{t=0}^{T} \left( \int_{[a,b]^{2}} \text{d}x_{1}^{t} \text{d}x_{2}^{t} f_{i}^{t+1} \left( x_{i}^{t+1}, x_{i}^{t}, x_{j}^{t}, x_{1}^{t}, x_{2}^{t} \right) \right. \times \\
    &\times \left. \sum_{\alpha_{1}^{t}} \sum_{\alpha_{i}^{t}} A_{1 i}^{t} \left[ \alpha_{1}^{t}, \alpha_{i}^{t} \right] u_{\alpha_{1}^{t}} \left( x_{1}^{t} \right) u_{\alpha_{i}^{t}} \left(x_{i}^{t}\right) \otimes \sum_{\beta_{2}^{t}} \sum_{\beta_{i}^{t}} A_{2 i}^{t} \left[ \beta_{2}^{t}, \beta_{i}^{t} \right] u_{\beta_{2}^{t}} \left(x_{2}^{t}\right) u_{\beta_{i}^{t}} \left( x_{i}^{t} \right) \right)\\
    =& \prod_{t=0}^{T} \left( \sum_{\alpha_{1}^{t}, \beta_{2}^{t}} \sum_{\alpha_{i}^{t}, \beta_{i}^{t}} A_{1 i}^{t} \left[ \alpha_{1}^{t}, \alpha_{i}^{t} \right] \otimes A_{2 i}^{t} \left[ \beta_{2}^{t}, \beta_{i}^{t} \right] u_{\alpha_{i}^{t}} \left(x_{i}^{t}\right) u_{\beta_{i}^{t}} \left(x_{i}^{t}\right) \right. \times \\
    &\times \left. \int_{[a,b]^{2}} \text{d}x_{1}^{t} \text{d}x_{2}^{t} f_{i}^{t+1} \left(x_{i}^{t+1}, x_{i}^{t}, x_{j}^{t}, x_{1}^{t}, x_{2}^{t}\right) u_{\alpha_{1}^{t}} \left(x_{1}^{t}\right) u_{\beta_{2}^{t}} \left(x_{2}^{t}\right) \right)\\
\intertext{Decomposing $f_{i}^{t+1} \left(x_{i}^{t+1}, x_{i}^{t}, x_{j}^{t}, x_{1}^{t}, x_{2}^{t}\right)$ into the basis:}
    =& \prod_{t=0}^{T} \sum_{\alpha_{1}^{t}, \beta_{2}^{t}} \sum_{\alpha_{i}^{t}, \beta_{i}^{t}} A_{1 i}^{t} \left[ \alpha_{1}^{t}, \alpha_{i}^{t}\right] \otimes A_{2 i}^{t} \left[ \beta_{2}^{t}, \beta_{i}^{t} \right] u_{\alpha_{i}^{t}} \left(x_{i}^{t}\right) u_{\beta_{i}^{t}} \left(x_{i}^{t}\right) \times\\
    &\times \int_{[a,b]^{2}} \text{d}x_{1}^{t} \text{d}x_{2}^{t} \sum_{\eta_{i}^{t+1}, \gamma_{i}^{t}, \gamma_{j}^{t}, \gamma_{1}^{t}, \gamma_{2}^{t}} F_{i}^{t+1} \left[\eta_{i}^{t+1}, \gamma_{i}^{t}, \gamma_{j}^{t}, \gamma_{1}^{t}, \gamma_{2}^{t}\right] \times\\
    & \times u_{\eta_{i}^{t+1}} \left(x_{i}^{t+1}\right) u_{\gamma_{i}^{t}} \left(x_{i}^{t}\right) u_{\gamma_{j}^{t}} \left(x_{j}^{t}\right) u_{\gamma_{1}^{t}} \left(x_{1}^{t}\right) u_{\gamma_{2}^{t}} \left(x_{2}^{t}\right) u_{\alpha_{1}^{t}} \left(x_{1}^{t}\right) u_{\beta_{2}^{t}} \left(x_{2}^{t}\right)\\
    =& \prod_{t=0}^{T} \left( \sum_{\alpha_{1}^{t}, \beta_{2}^{t}} \sum_{\alpha_{i}^{t}, \beta_{i}^{t}} A_{1 i}^{t} \left[ \alpha_{1}^{t}, \alpha_{i}^{t} \right] \otimes A_{2 i}^{t} \left[ \beta_{2}^{t}, \beta_{i}^{t}\right] u_{\alpha_{i}^{t}} \left(x_{i}^{t}\right) u_{\beta_{i}^{t}} \left(x_{i}^{t}\right) \right. \times \\
    & \times \left. \sum_{\eta_{i}^{t+1}, \gamma_{i}^{t}, \gamma_{j}^{t}} F_{i}^{t+1} \left[\eta_{i}^{t+1}, \gamma_{i}^{t}, \gamma_{j}^{t}, -\alpha_{1}^{t}, -\alpha_{2}^{t} \right] u_{\eta_{i}^{t+1}} \left(x_{i}^{t+1}\right) u_{\gamma_{i}^{t}} \left(x_{i}^{t}\right)u_{\gamma_{j}^{t}} \left(x_{j}^{t}\right)\right)\\
    =& \sum_{\overline{\eta}_{i}} \prod_{t=0}^{T} \left( \sum_{\alpha_{1}^{t}, \beta_{2}^{t}} \sum_{\alpha_{i}^{t}, \beta_{i}^{t}} A_{1 i}^{t} \left[ \alpha_{1}^{t}, \alpha_{i}^{t} \right] \otimes A_{2 i}^{t} \left[ \beta_{2}^{t}, \beta_{i}^{t} \right] u_{\alpha_{i}^{t}} \left(x_{i}^{t}\right) u_{\beta_{i}^{t}} \left(x_{i}^{t}\right) \right. \times \\
    & \times \left. \sum_{\gamma_{i}^{t}, \gamma_{j}^{t}} F_{i}^{t+1} \left[ \eta_{i}^{t+1}, \gamma_{i}^{t}, \gamma_{j}^{t}, -\alpha_{1}^{t}, -\alpha_{2}^{t} \right] u_{\eta_{i}^{t+1}} \left(x_{i}^{t+1}\right) u_{\gamma_{i}^{t}} \left(x_{i}^{t}\right) u_{\gamma_{j}^{t}}\left(x_{j}^{t}\right) \right)
\intertext{Now one can re-associate each of the $u_{\eta_{i}^{t+1}} \left(x_{i}^{t+1}\right)$ terms with the successive time:}
    =&\sum_{\overline{\eta}_{i}} \prod_{t=0}^{T} \left( \sum_{\alpha_{1}^{t}, \beta_{2}^{t}} \sum_{\alpha_{i}^{t}, \beta_{i}^{t}} A_{1 i}^{t} \left[\alpha_{1}^{t}, \alpha_{i}^{t}\right] \otimes A_{2 i}^{t} \left[\beta_{2}^{t}, \beta_{i}^{t}\right] u_{\alpha_{i}^{t}} \left(x_{i}^{t}\right) u_{\beta_{i}^{t}} \left(x_{i}^{t}\right) \right. \times \\
    &\times \left. \sum_{\gamma_{i}^{t}, \gamma_{j}^{t}} F_{i}^{t+1} \left[\eta_{i}^{t+1}, \gamma_{i}^{t}, \gamma_{j}^{t}, -\alpha_{1}^{t}, -\alpha_{2}^{t}\right] u_{\eta_{i}^{t}} \left(x_{i}^{t}\right) u_{\gamma_{i}^{t}} \left(x_{i}^{t}\right) u_{\gamma_{j}^{t}} \left(x_{j}^{t}\right) \right)
\intertext{And finally, the whole matrix-product state can be projected onto the basis $\left\{ u_{\delta_i^t}(x) \right\}_{t}$}
    =& \sum_{\overline{\eta}_{i}} \prod_{t=0}^{T} \sum_{\delta_{i}^{t}} \int_{\left[a,b\right]} \text{d}y_{i}^{t} \left( \sum_{\alpha_{1}^{t}, \beta_{2}^{t}} \sum_{\alpha_{i}^{t}, \beta_{i}^{t}} A_{1 i}^{t} \left[\alpha_{1}^{t}, \alpha_{i}^{t}\right] \otimes A_{2 i}^{t} \left[\beta_{2}^{t}, \beta_{i}^{t}\right] u_{\alpha_{i}^{t}} \left(y_{i}^{t}\right) u_{\beta_{i}^{t}} \left(y_{i}^{t}\right) \right. \times \\
    & \times \left. \sum_{\gamma_{i}^{t}, \gamma_{j}^{t}} F_{i}^{t+1} \left[\eta_{i}^{t+1}, \gamma_{i}^{t}, \gamma_{j}^{t}, -\alpha_{1}^{t}, -\alpha_{2}^{t} \right] u_{\eta_{i}^{t}} \left(y_{i}^{t}\right) u_{\gamma_{i}^{t}} \left(y_{i}^{t}\right) u_{\gamma_{j}^{t}} \left(x_{j}^{t}\right) u_{\delta_{i}^{t}}^{*} \left(y_{i}^{t}\right) \right) u_{\delta_{i}^{t}} \left(x_{i}^{t}\right) \\
    =& \sum_{\overline{\eta}_{i}} \prod_{t=0}^{T} \sum_{\alpha_{1}^{t}, \beta_{2}^{t}} \sum_{\alpha_{i}^{t}, \beta_{i}^{t}} A_{1 i}^{t} \left[\alpha_{1}^{t}, \alpha_{i}^{t}\right] \otimes A_{2 i}^{t} \left[\beta_{2}^{t}, \beta_{i}^{t}\right] \sum_{\gamma_{i}^{t}, \gamma_{j}^{t}} F_{i}^{t+1} \left[\eta_{i}^{t+1}, \gamma_{i}^{t}, \gamma_{j}^{t}, -\alpha_{1}^{t}, -\alpha_{2}^{t} \right] \times \\
    & \times \sum_{\delta_{i}^{t}} \int_{\left[a,b\right]} \text{d}y_{i}^{t} \left[ u_{\alpha_{i}^{t}} \left(y_{i}^{t}\right) u_{\beta_{i}^{t}} \left(y_{i}^{t}\right) u_{\eta_{i}^{t}} \left(y_{i}^{t}\right) u_{\gamma_{i}^{t}} \left(y_{i}^{t}\right) u_{\delta_{i}^{t}}^{*} \left(y_{i}^{t}\right) \right] u_{\delta_{i}^{t}} \left(x_{i}^{t}\right) u_{\gamma_{j}^{t}} \left(x_{j}^{t}\right) \\
    =& \sum_{\overline{\eta}_{i}} \prod_{t=0}^{T} \left[ \left( \sum_{\alpha_{1}^{t}, \beta_{2}^{t}} \sum_{\alpha_{i}^{t}, \beta_{i}^{t}} A_{1 i}^{t} \left[\alpha_{1}^{t}, \alpha_{i}^{t}\right] \otimes A_{2 i}^{t} \left[\beta_{2}^{t}, \beta_{i}^{t}\right] \sum_{\gamma_{i}^{t}, \gamma_{j}^{t}} F_{i}^{t+1} \left[\eta_{i}^{t+1}, \gamma_{i}^{t}, \gamma_{j}^{t}, -\alpha_{1}^{t}, -\alpha_{2}^{t} \right] \right. \right. \times \\
    & \times \left. \left. \sum_{\delta_{i}^{t}} \Gamma_{i}^{t} \left[\alpha_{i}^{t}, \beta_{i}^{t}, \gamma_{i}^{t}, \delta_{i}^{t}, \eta_{i}^{t} \right] \right) u_{\delta_{i}^{t}} \left(x_{i}^{t}\right) u_{\gamma_{j}^{t}} \left(x_{j}^{t}\right) \right] \\
    =& \sum_{\overline{\eta}_{i}, \overline{\delta}_i, \overline{\gamma}_j} \left(\prod_{t=0}^{T} B_{ij}^{t} [\eta_{i}^{t+1}, \eta_{i}^{t}, \delta_{i}^{t}, \gamma_{j}^{t}] \right) \left( \prod_{t=0}^{T} u_{\delta_{i}^{t}} \left(x_{i}^{t}\right) u_{\gamma_{j}^{t}} \left(x_{j}^{t}\right) \right)
\end{align*}
having defined
\begin{equation*}
B_{ij}^{t} [\eta_{i}^{t+1}, \eta_{i}^{t}, \delta_{i}^{t}, \gamma_{j}^{t}] \coloneqq \sum_{\alpha_{1}^{t}, \beta_{2}^{t}} \sum_{\alpha_{i}^{t}, \beta_{i}^{t}} A_{1 i}^{t} \left[\alpha_{1}^{t}, \alpha_{i}^{t}\right] \otimes A_{2 i}^{t} \left[\beta_{2}^{t}, \beta_{i}^{t}\right] \sum_{\gamma_{i}^{t}} F_{i}^{t+1} \left[\eta_{i}^{t+1}, \gamma_{i}^{t}, \gamma_{j}^{t}, -\alpha_{1}^{t}, -\alpha_{2}^{t} \right] \Gamma_{i}^{t} \left[\alpha_{i}^{t}, \beta_{i}^{t}, \gamma_{i}^{t}, \delta_{i}^{t}, \eta_{i}^{t} \right].
\end{equation*}
This is almost in the desired final form; indeed, one wants the matrices $B_{ij}^{t}$ to depend on only one $\eta_i^t$. As the matrix-product state $B_{ij}$ for the basis expansion coefficients is discrete, this can be done by a sweep of SVD, in a completely analogous way to \cite{barthelMatrixProductAlgorithm2018,crottiMatrixProductBelief2023c}.\\
SVD is a factorization of an arbitrary matrix $M\in \mathbb{C}^{m \times n}$ of the form $M = U S V$, with $U \in \mathbb{C}^{m \times m}$ unitary, $V \in \mathbb{C}^{n \times n}$ unitary, and $S \in \mathbb{R}^{m \times n}$ diagonal with non-negative entries on the diagonal, called the singular values of $M$. In this case, it can be used to separate the indices, as explained in \Cref{alg:svd-sweep} (the last term does not depend on $\eta_i^{T+1}$ by definition, hence the algorithm is coherent).
\begin{tcolorbox}[float*=htbp, width=\textwidth]
\setlength{\algoheightrule}{0pt}
\setlength{\algotitleheightrule}{0pt}
\begin{algorithm}[H]
    \caption{Sweep of SVD\\}
    \label{alg:svd-sweep}
    \SetAlgoLined
    \DontPrintSemicolon
    
    \For{$t \in 0:T$}{
        - Define $D_{ij}^{t}$ as $\left[ D_{ij}^{t} \right]_{\left( m^t, \eta_{i}^{t} \delta_{i}^{t}, \gamma_{j}^{t} \right), \left( m^{t+1}, \eta_{i}^{t+1} \right)} \coloneqq \left[ B_{ij}^{t} [\eta_{i}^{t+1}, \eta_{i}^{t}, \delta_{i}^{t}, \gamma_{j}^{t}] \right]_{m^t, m^{t+1}}$\;
        - Perform SVD on $D_{ij}^t$, obtaining $\left[ D_{ij}^{t} \right]_{\left( m^t, \eta_{i}^{t}, \delta_{i}^{t}, \gamma_{j}^{t} \right), \left( m^{t+1}, \eta_{i}^{t+1} \right)} \stackrel{\text{SVD}}{=} \sum_{k} [U^t]_{\left( m^t, \eta_{i}^{t}, \delta_{i}^{t}, \gamma_{j}^{t} \right), k} [M^t]_{k,k} [V^t]_{k, \left( m^{t+1}, \eta_{i}^{t+1} \right)}$\;
        - Define $\left[ \tilde{C}_{ij}^{t} [\eta_i^t, \delta_i^t, \gamma_j^t] \right]_{m^t, k} = [U^t]_{\left( m^t, \eta_{i}^{t}, \delta_{i}^{t}, \gamma_{j}^{t} \right), k}$ (therefore identifying $m^{t+1}$ with k)\;
        - Redefine $\left[ B_{ij}^{t+1} \left[ \eta_{i}^{t+2}, \eta_{i}^{t+1}, \delta_{i}^{t+1}, \gamma_{j}^{t+1} \right] \right]_{k, m^{t+2}} = \sum_{m^{t+1}} [M^t]_{k,k} [V^t]_{k, \left( m^{t+1}, \eta_{i}^{t+1}\right)} \left[ B_{ij}^{t+1} [\eta_{i}^{t+2}, \eta_{i}^{t+1}, \delta_{i}^{t+1}, \gamma_{j}^{t+1}] \right]_{m^{t+1}, m^{t+2}}$
        }
\end{algorithm}
\end{tcolorbox}

At this point, it is easy to close the matrix-product \textit{ansatz}:
\begin{align*}
    \mu_{ij} \left( \overline{x}_{i}, \overline{x}_{j} \right) =& \sum_{\overline{\eta}_{i}, \overline{\delta}_i, \overline{\gamma}_j} \left(\prod_{t=0}^{T} B_{ij}^{t} [\eta_{i}^{t+1}, \eta_{i}^{t}, \delta_{i}^{t}, \gamma_{j}^{t}] \right) \left( \prod_{t=0}^{T} u_{\delta_{i}^{t}} \left(x_{i}^{t}\right) u_{\gamma_{j}^{t}} \left(x_{j}^{t}\right) \right) \\
    =& \sum_{\overline{\delta}_i, \overline{\gamma}_j} \left(\prod_{t=0}^{T} \sum_{\eta_i^t} \tilde{C}_{ij}^{t} [\eta_{i}^{t}, \delta_{i}^{t}, \gamma_{j}^{t}] \right) \left( \prod_{t=0}^{T} u_{\delta_{i}^{t}} \left(x_{i}^{t}\right) u_{\gamma_{j}^{t}} \left(x_{j}^{t}\right) \right) \\
    =& \prod_{t=0}^{T} \sum_{\delta_i^t, \gamma_j^t} C_{ij}^{t} [\delta_{i}^{t}, \gamma_{j}^{t}] u_{\delta_{i}^{t}} \left(x_{i}^{t}\right) u_{\gamma_{j}^{t}} \left(x_{j}^{t}\right)
\end{align*}
with $C_{ij}^{t} [\delta_{i}^{t}, \gamma_{j}^{t}] \coloneqq \sum_{\eta_i^t} \tilde{C}_{ij}^{t} [\eta_{i}^{t}, \delta_{i}^{t}, \gamma_{j}^{t}]$.\\
From the above derivation, it can be noted that the dimensions of the matrices of the matrix-product state grow during the iteration, as a result of the performed Kronecker products. For this reason, it is convenient to perform a second sweep of SVD, this time starting from $t=T$ and going backwards towards $t=0$. During this sweep, just some singular values are retained (the $\sum_k$ is performed only on the values of $k$ that yield the largest values of $M_{k,k}$), providing a controlled approximation of the matrix-product state. Further explanation on the truncation algorithm can be found in \cite{oseledetsTensorTrainDecomposition2011a}, while the details on the accuracy of the truncation can be found in \cref{app:bound-SVD}.

\section{MPBP step in alternative factor graph}
\label{app:alternative-factor-graph-calculation}
Here, the detailed calculation is shown for an MPBP update step in the factor graph in \cref{fig:alternative-factor-graph}.\\
Given the tree structure of the subgraph under consideration, the update will be performed in the most efficient way, so as to require the minimum number of computations.
\begin{align*}
    \mu_{a b} \left( \overline{x}_{i}, \overline{y}_{12} \right) =& \sum_{\overline{x}_{1}, \overline{x}_{2}} \prob{\overline{y}_{12} | \overline{x}_{1}, \overline{x}_{2}} \mu_{1 i} \left( \overline{x}_{1}, \overline{x}_{i} \right) \mu_{2 i} \left( \overline{x}_{2}, \overline{x}_{i} \right) \\[0.5cm]
    \mu_{b c} \left( \overline{x}_{i}, \overline{y}_{123} \right) =& \sum_{\overline{x}_{3}, \overline{y}_{12}} \prob{ \overline{y}_{123} | \overline{x}_{3}, \overline{y}_{12}} \mu_{3 i} \left( \overline{x}_{3}, \overline{x}_{i} \right) \mu_{a b} \left( \overline{x}_{i}, \overline{y}_{12} \right)\\
    =& \sum_{\overline{x}_{3}, \overline{y}_{12}} \prob{ \overline{y}_{123} | \overline{x}_{3}, \overline{y}_{12}} \mu_{3 i} \left( \overline{x}_{3}, \overline{x}_{i} \right) \sum_{\overline{x}_{1}, \overline{x}_{2}} \prob{\overline{y}_{12} | \overline{x}_{1}, \overline{x}_{2}} \mu_{1 i} \left( \overline{x}_{1}, \overline{x}_{i} \right) \mu_{2 i} \left( \overline{x}_{2}, \overline{x}_{i} \right)\\
    =& \sum_{\overline{x}_{1}, \overline{x}_{2}, \overline{x}_{3}} \prob{ \overline{y}_{123} | \overline{x}_{1}, \overline{x}_{2}, \overline{x}_{3}} \mu_{1 i} \left( \overline{x}_{1}, \overline{x}_{i} \right) \mu_{2 i} \left( \overline{x}_{2}, \overline{x}_{i} \right) \mu_{3 i} \left( \overline{x}_{3}, \overline{x}_{i} \right)\\[0.5cm]
    \mu_{c i} \left( \overline{x}_{i}, \overline{y}_{123j} \right) =& \sum_{\overline{x}_{j}, \overline{y}_{123}} \prob{\overline{y}_{123j} | \overline{x}_{j}, \overline{y}_{123}} \mu_{ji} \left( \overline{x}_{j}, \overline{x}_{i} \right) \mu_{b c} \left( \overline{x}_{i}, \overline{y}_{123} \right)\\
    =& \sum_{\overline{x}_{j}, \overline{y}_{123}} \prob{\overline{y}_{123j} | \overline{x}_{j}, \overline{y}_{123}} \mu_{ji} \left( \overline{x}_{j}, \overline{x}_{i} \right) \sum_{\overline{x}_{1}, \overline{x}_{2}, \overline{x}_{3}} \prob{\overline{y}_{123}, \overline{x}_{1}, \overline{x}_{2}, \overline{x}_{3}} \mu_{1 i} \left( \overline{x}_{1}, \overline{x}_{i} \right) \mu_{2 i} \left( \overline{x}_{2}, \overline{x}_{i} \right) \mu_{3 i} \left( \overline{x}_{3}, \overline{x}_{i} \right)\\
    =& \sum_{\overline{x}_{1}, \overline{x}_{2}, \overline{x}_{3}, \overline{x}_{j}} \prob{ \overline{y}_{123j} | \overline{x}_{1}, \overline{x}_{2}, \overline{x}_{3}, \overline{x}_{j}} \mu_{1 i} \left( \overline{x}_{1}, \overline{x}_{i} \right) \mu_{2 i} \left( \overline{x}_{2}, \overline{x}_{i} \right) \mu_{3 i} \left( \overline{x}_{3}, \overline{x}_{i} \right) \mu_{ji} \left( \overline{x}_{j}, \overline{x}_{i} \right)\\[0.5cm]
    \mu_{i c} \left( \overline{x}_{i}, \overline{y}_{123j} \right) =& \;\; \Phi_i\left( \overline{x}_{i}, \{\overline{y}_{123j}\} \right) \\[0.5cm]
    \mu_{ij} \left( \overline{x}_{i}, \overline{x}_{j} \right) =& \sum_{\overline{y}_{123}, \overline{y}_{123j}} \prob{ \overline{y}_{123j} | \overline{x}_{j}, \overline{y}_{123}} \mu_{i c} \left( \overline{x}_{i}, \overline{y}_{123j} \right) \mu_{b c} \left( \overline{x}_{i}, \overline{y}_{123} \right) \\
    =& \sum_{\overline{y}_{123}, \overline{y}_{123j}} \prob{ \overline{y}_{123j} | \overline{x}_{j}, \overline{y}_{123}} \Phi_i\left( \overline{x}_{i}, \{\overline{y}_{123j}\} \right) \sum_{\overline{x}_{1}, \overline{x}_{2}, \overline{x}_{3}} \prob{ \overline{y}_{123} | \overline{x}_{1}, \overline{x}_{2}, \overline{x}_{3}} \mu_{1 i} \left( \overline{x}_{1}, \overline{x}_{i} \right) \mu_{2 i} \left( \overline{x}_{2}, \overline{x}_{i} \right) \mu_{3 i} \left( \overline{x}_{3}, \overline{x}_{i} \right)\\
    =& \sum_{\overline{x}_{1}, \overline{x}_{2}, \overline{x}_{3}} \Phi_i\left( \overline{x}_{i}, \{\overline{x}_{1}, \overline{x}_{2}, \overline{x}_{3}, \overline{x}_{j}\} \right) \mu_{1 i} \left( \overline{x}_{1}, \overline{x}_{i} \right) \mu_{2 i} \left( \overline{x}_{2}, \overline{x}_{i} \right) \mu_{3 i} \left( \overline{x}_{3}, \overline{x}_{i} \right) \\[0.5cm]
    \mu_{c b} \left( \overline{x}_{i}, \overline{y}_{123} \right) =& \sum_{\overline{x}_{j}, \overline{y}_{123j}} \prob{ \overline{y}_{123j} | \overline{x}_{j}, \overline{y}_{123}} \mu_{ji} \left( \overline{x}_{j}, \overline{x}_{i} \right) \mu_{i c} \left( \overline{x}_{i}, \overline{y}_{123j} \right)\\
    =& \sum_{\overline{x}_{j}, \overline{y}_{123j}} \prob{ \overline{y}_{123j} | \overline{x}_{j} + \overline{y}_{123}} \Phi_i\left( \overline{x}_{i}, \{\overline{y}_{123j}\} \right) \mu_{ji} \left( \overline{x}_{j}, \overline{x}_{i} \right) \\
    =& \sum_{\overline{x}_{j}} \Phi_i\left( \overline{x}_{i}, \{\overline{x}_{j}, \overline{y}_{123}\} \right) \mu_{ji} \left( \overline{x}_{j}, \overline{x}_{i} \right) \\[0.5cm]
    \mu_{i 3} \left( \overline{x}_{i}, \overline{x}_{3} \right) =& \sum_{\overline{y}_{123}, \overline{y}_{12}} \prob{ \overline{y}_{123} | \overline{x}_{3}, \overline{y}_{12}} \mu_{c b} \left( \overline{x}_{i}, \overline{y}_{123} \right) \mu_{a b} \left( \overline{x}_{i}, \overline{y}_{12} \right) \\
    =& \sum_{\overline{y}_{123}, \overline{y}_{12}} \prob{ \overline{y}_{123} | \overline{x}_{3}, \overline{y}_{12}} \sum_{\overline{x}_{j}} \Phi_i\left( \overline{x}_{i}, \{\overline{x}_{j}, \overline{y}_{123}\} \right) \mu_{ji} \left( \overline{x}_{j}, \overline{x}_{i} \right) \sum_{\overline{x}_{1}, \overline{x}_{2}} \prob{\overline{y}_{12} | \overline{x}_{1}, \overline{x}_{2}} \mu_{1 i} \left( \overline{x}_{1}, \overline{x}_{i} \right) \mu_{2 i} \left( \overline{x}_{2}, \overline{x}_{i} \right) \\
    =& \sum_{\overline{x}_{1}, \overline{x}_{2}, \overline{x}_{j}} \Phi_i\left( \overline{x}_{i}, \{\overline{x}_{1}, \overline{x}_{2}, \overline{x}_{3}, \overline{x}_{j}\} \right) \mu_{1 i} \left( \overline{x}_{1}, \overline{x}_{i} \right) \mu_{2 i} \left( \overline{x}_{2}, \overline{x}_{i} \right) \mu_{ji} \left( \overline{x}_{j}, \overline{x}_{i} \right) \\[0.5cm]
    \mu_{b a} \left( \overline{x}_{i}, \overline{y}_{12} \right) =& \sum_{\overline{x}_{3}, \overline{y}_{123}} \prob{ \overline{y}_{123} | \overline{x}_{3}, \overline{y}_{12}} \mu_{3 i} \left( \overline{x}_{3}, \overline{x}_{i} \right) \mu_{c b} \left( \overline{x}_{i}, \overline{y}_{123} \right)\\
    =& \sum_{\overline{x}_{3}, \overline{y}_{123}} \prob{ \overline{y}_{123} | \overline{x}_{3}, \overline{y}_{12}} \mu_{3 i} \left( \overline{x}_{3}, \overline{x}_{i} \right) \sum_{\overline{x}_{j}} \Phi_i\left( \overline{x}_{i}, \{\overline{x}_{j}, \overline{y}_{123}\} \right) \mu_{ji} \left( \overline{x}_{j}, \overline{x}_{i} \right)\\
    =& \sum_{\overline{x}_{3}, \overline{x}_{j}} \Phi_i\left( \overline{x}_{i}, \{\overline{x}_{3}, \overline{x}_{j}, \overline{y}_{12}\} \right) \mu_{3 i} \left( \overline{x}_{3}, \overline{x}_{i} \right) \mu_{ji} \left( \overline{x}_{j}, \overline{x}_{i} \right) \\[0.5cm]
    \mu_{i 2} \left( \overline{x}_{i}, \overline{x}_{2} \right) =& \sum_{\overline{x}_{1}, \overline{y}_{12}} \prob{ \overline{y}_{12} | \overline{x}_{1}, \overline{x}_{2}} \mu_{1 i} \left( \overline{x}_{1}, \overline{x}_{i} \right) \mu_{b a} \left( \overline{x}_{i}, \overline{y}_{12} \right) \\
    =& \sum_{\overline{x}_{1}, \overline{y}_{12}} \prob{ \overline{y}_{12} | \overline{x}_{1}, \overline{x}_{2}} \mu_{1 i} \left( \overline{x}_{1}, \overline{x}_{i} \right) \sum_{\overline{x}_{3}, \overline{x}_{j}} \Phi_i\left( \overline{x}_{i}, \{\overline{x}_{3}, \overline{x}_{j}, \overline{y}_{12}\} \right) \mu_{3 i} \left( \overline{x}_{3}, \overline{x}_{i} \right) \mu_{ji} \left( \overline{x}_{j}, \overline{x}_{i} \right)\\
    =& \sum_{\overline{x}_{1}, \overline{x}_{3}, \overline{x}_{j}} \Phi_i\left( \overline{x}_{i}, \{\overline{x}_{1}, \overline{x}_{2}, \overline{x}_{3}, \overline{x}_{j}\} \right) \mu_{1 i} \left( \overline{x}_{1}, \overline{x}_{i} \right) \mu_{3 i} \left( \overline{x}_{3}, \overline{x}_{i} \right) \mu_{ji} \left( \overline{x}_{j}, \overline{x}_{i} \right) \\[0.5cm]
    \mu_{i 1} \left( \overline{x}_{i}, \overline{x}_{1} \right) =& \sum_{\overline{x}_{2}, \overline{y}_{12}} \prob{ \overline{y}_{12} | \overline{x}_{1}, \overline{x}_{2}} \mu_{2 i} \left( \overline{x}_{2}, \overline{x}_{i} \right) \mu_{b a} \left( \overline{x}_{i}, \overline{y}_{12} \right) \\
    =& \sum_{\overline{x}_{2}, \overline{y}_{12}} \prob{ \overline{y}_{12} | \overline{x}_{1}, \overline{x}_{2}} \mu_{2 i} \left( \overline{x}_{2}, \overline{x}_{i} \right) \sum_{\overline{x}_{3}, \overline{x}_{j}} \Phi_i\left( \overline{x}_{i}, \{\overline{x}_{3}, \overline{x}_{j}, \overline{y}_{12}\} \right) \mu_{3 i} \left( \overline{x}_{3}, \overline{x}_{i} \right) \mu_{ji} \left( \overline{x}_{j}, \overline{x}_{i} \right)\\
    =& \sum_{\overline{x}_{2}, \overline{x}_{3}, \overline{x}_{j}} \Phi_i\left( \overline{x}_{i}, \{\overline{x}_{1}, \overline{x}_{2}, \overline{x}_{3}, \overline{x}_{j}\} \right) \mu_{2 i} \left( \overline{x}_{2}, \overline{x}_{i} \right) \mu_{3 i} \left( \overline{x}_{3}, \overline{x}_{i} \right) \mu_{ji} \left( \overline{x}_{j}, \overline{x}_{i} \right)
\end{align*}
Here, the identity $\sum_{\overline{y}} \prob{\overline{y} | \overline{x}_1, \overline{x}_2} \Phi\left(\overline{z}, \{\overline{y}\} \right) = \Phi\left(\overline{z}, \{\overline{x}_1, \overline{x}_2\} \right)$ has been applied repeatedly. This is much better understood when $\Phi$ has the typical form of a non-reweighted model:
\begin{align*}
    \sum_{\overline{y}} \prob{\overline{y} | \overline{x}_1, \overline{x}_2} \Phi\left(\overline{z}, \{\overline{y}\} \right) =& \sum_{\overline{y}} \prod_t \prob{y^t | x_1^t, x_2^t} \prod_t \prob{ z^{t+1} | \{y^t\}, x_t} \\
    =& \prod_t \sum_{y^t} \prob{ z^{t+1} | \{y^t\}, x_t} \prob{y^t | x_1^t, x_2^t} =\\
    =& \prod_t \prob{ z^{t+1} | \{x_1^t, x_2^t\}, x_t} \\
    =& \;\; \Phi\left(\overline{z}, \{\overline{x}_1, \overline{x}_2\} \right).
\end{align*}
It is also worth noticing that, in computing the outgoing messages $\mu_{i 1}$, $\mu_{i 2}$, $\mu_{i 3}$ and $\mu_{ij}$, one sweep of SVD is required for each message to obtain the same form as the incoming messages.   

\section{Truncation}
\label{app:bound-SVD}
Similarly to what happens in the discrete case without the basis expansion, SVD truncation (sometimes called compression) is controlled and can be rendered as accurate as needed.  
Take a matrix-product state with $T$ matrices expressed in a finite basis expansion of $M$ elements
\begin{equation*}
    f(\overline{x}) = \prod_t A^t(x^t) = \prod_t \sum_{\alpha^t=1}^M \hat{A}^t[\alpha^t] u_{\alpha^t}(x^t) = \sum_{\overline{\alpha}} \hat{f}[\overline{\alpha}] \prod_t u_{\alpha^t}(x^t)
\end{equation*}
where $\hat{f}[\overline{\alpha}] \coloneqq \prod_t \hat{A}^t[\alpha^t]$ is the matrix-product state of the coefficients of the basis expansion.\\
Then, obtain $\hat{g}$ as the $\varepsilon$-truncated version of $\hat{f}$ (reducing bond dimension from $d$ to $d'$) with the SVD-based truncation described in \cite{oseledetsTensorTrainDecomposition2011a} so that 
\begin{equation*}
    \left|\left| \hat{f} - \hat{g} \right|\right|_F^2 = \sum_{\overline{\alpha}} |\hat{f}[\overline{\alpha}]-\hat{g}[\overline{\alpha}]|^2\leq \varepsilon
\end{equation*}
and define the truncated version of $f$ as
\begin{equation*}
    g(\overline{x}) = \sum_{\overline{\alpha}} \hat{g}[\overline{\alpha}] \prod_t u_{\alpha^t}(x^t).
\end{equation*}

One gets:
\begin{align*}
    \left|\left| f - g \right|\right|^2 =& \int \diff{\overline{x}} \left| f(\overline{x}) - g(\overline{x}) \right|^2 \\
    =& \int \diff{\overline{x}} \left| \sum_{\overline{\alpha}} \left( \hat{f}[\overline{\alpha}] - \hat{g}[\overline{\alpha}] \right) \prod_t u_{\alpha^t}(x^t) \right|^2 \\
    =& \int \diff{\overline{x}} \left( \sum_{\overline{\alpha}} \left( \hat{f}[\overline{\alpha}] - \hat{g}[\overline{\alpha}] \right) \prod_t u_{\alpha^t}(x^t) \right) \left( \sum_{\overline{\beta}} \left( \hat{f}[\overline{\beta}] - \hat{g}[\overline{\beta}] \right)^* \prod_t u_{\beta^t}^*(x^t) \right) \\
    =& \sum_{\overline{\alpha}, \overline{\beta}} \left( \hat{f}[\overline{\alpha}] - \hat{g}[\overline{\alpha}] \right) \left( \hat{f}[\overline{\beta}] - \hat{g}[\overline{\beta}] \right)^* \prod_t \int \diff{x^t} u_{\alpha^t}(x^t) u_{\beta^t}^*(x^t) \\
    =& \sum_{\overline{\alpha}, \overline{\beta}} \left( \hat{f}[\overline{\alpha}] - \hat{g}[\overline{\alpha}] \right) \left( \hat{f}[\overline{\beta}] - \hat{g}[\overline{\beta}] \right)^* \prod_t \delta_{\alpha^t, \beta^t} \\
    =& \sum_{\overline{\alpha}} \left| \hat{f}[\overline{\alpha}] - \hat{g}[\overline{\alpha}] \right|^2 \\
    \leq&~\varepsilon.
\end{align*}
Therefore, the approximation is controlled and can be made as accurate as desired by retaining matrices with large enough bond dimensions in the approximation of the matrix-product state of the coefficients.

\section{Calculations for convolution in the Fourier Basis}
\label{app:convolution}
Below are the detailed calculations for the convolution of messages written in the Fourier basis, introduced in \cref{sec:convolution}.
\begin{align*}
    \mu_{A\cup B, i} (\overline{y}_{A\cup B}, \overline{x}_i) =& \int \diff{\overline{y}_A} \diff{\overline{y}_B} \delta(\overline{y}_{A\cup B} - \overline{y}_A - \overline{y}_B) \mu_{A i} (\overline{y}_A, \overline{x}_i) \mu_{B i} (\overline{y}_B, \overline{x}_i)\\
    =& \prod_t \int \diff{y_A^t} \diff{y_B^t} \delta(y_{A\cup B}^t - y_A^t - y_B^t) A_{A i}^t (y_A^t, x_i^t) \otimes A_{B i}^t (y_B^t, x_i^t)\\
    =& \sum_{\overline{m}_A, \overline{m}_B} \prod_t \int \diff{y_A^t} \diff{y_B^t} \delta(y_{A\cup B}^t - y_A^t - y_B^t) \left[ A_{A i}^t (y_A^t, x_i^t) \right]_{m_A^t, m_A^{t+1}} \left[ A_{B i}^t (y_B^t, x_i^t) \right]_{m_B^t, m_B^t+1}.
\end{align*}
By writing $\mu_{A\cup B, i} (\overline{y}_{A\cup B}, \overline{x}_i) = \sum_{\overline{n}} \prod_t \left[ A_{A\cup B, i}^t(y_{A\cup B}^t, x_i^t) \right]_{n^t, n^{t+1}}$ and identifying the index $n^t$ with the multi-index $(m_A^t, m_B^t)$ one can write:
\begin{align*}
    \left[ A_{A\cup B, i}^t(y_{A\cup B}^t, x_i^t) \right]_{n^t, n^{t+1}} =& \int \diff{y_A^t} \diff{y_B^t} \delta(y_{A\cup B}^t - y_A^t - y_B^t) \times \\
    & \times \left( \sum_{\alpha_A^t} \left[ \hat{A}_{A i}^t (\alpha_A^t, x_i^t) \right]_{m_A^t, m_A^{t+1}} u_{\alpha_A^t}(y_A^t) \right) \left( \sum_{\alpha_B^t} \left[ \hat{A}_{B i}^t (\alpha_B^t, x_i^t) \right]_{m_B^t, m_B^t+1} u_{\alpha_B^t}(y_B^t) \right).
\end{align*}
The coefficients of the Fourier expansion can be calculated by projecting onto the basis elements:
\begin{align*}
    \left[ \hat{A}_{A\cup B, i}^t(\alpha_{A\cup B}^t, x_i^t) \right]_{n^t, n^{t+1}} =& \sum_{\alpha_A^t, \alpha_B^t} \left[ \hat{A}_{A i}^t (\alpha_A^t, x_i^t) \right]_{m_A^t, m_A^{t+1}} \left[ \hat{A}_{B i}^t (\alpha_B^t, x_i^t) \right]_{m_B^t, m_B^t+1}  \times \\
    &\times \int \diff{y_A^t} \diff{y_B^t} \diff{y_{A\cup B}^t} \delta(y_{A\cup B}^t - y_A^t - y_B^t) u_{\alpha_A^t}(y_A^t) u_{\alpha_B^t}(y_B^t) u_{\alpha_{A\cup B}^t}^*(y_{A\cup B}^t) \\
    =& \sum_{\alpha_A^t, \alpha_B^t} \left[ \hat{A}_{A i}^t (\alpha_A^t, x_i^t) \right]_{m_A^t, m_A^{t+1}} \left[ \hat{A}_{B i}^t (\alpha_B^t, x_i^t) \right]_{m_B^t, m_B^t+1}  \times \\
    &\times \int \diff{y_A^t} \diff{y_B^t} u_{\alpha_A^t}(y_A^t) u_{\alpha_B^t}(y_B^t) u_{\alpha_{A\cup B}^t}^*(y_A^t + y_B^t).
\end{align*}
Supposing that the functions of $y_A^t$ are $P_A$-periodic and the functions of $y_B^t$ are  $P_B$-periodic, the functions of $y_{A\cup B}^t$ can be defined to be $(P_A + P_B)$-periodic. The integral is thus evaluated as:
\begin{align*}
    I_{\alpha, \beta, \gamma} \coloneqq& \int \diff{y_A^t} \diff{y_B^t} u_{\alpha}(y_A^t) u_{\beta}(y_B^t) u_{\gamma}^*(y_A^t + y_B^t) \\
    =& \int_{-P_A/2}^{+P_A/2} \diff{y_A^t} e^{i\frac{2\pi}{P_A}\alpha y_A^t - i \frac{2\pi}{P_A + P_B} \gamma y_A^t} \int_{-P_B/2}^{+P_B/2} \diff{y_B^t} e^{i\frac{2\pi}{P_B}\beta y_B^t - i \frac{2\pi}{P_A + P_B} \gamma y_B^t} \\
    =& \left( -i \dfrac{P_A (P_A + P_B)}{2\pi (P_A+P_B)\alpha - P_A \gamma} \left[ e^{2\pi i \frac{(P_A+P_B)\alpha - P_A \gamma}{P_A (P_A + P_B)} y_A^t} \right]_{y_A^t = -P_A/2}^{y_A^t = +P_A/2} \right) \times\\
    &\times \left( -i \dfrac{P_B (P_A + P_B)}{2\pi (P_A+P_B)\beta - P_B \gamma} \left[ e^{2\pi i \frac{(P_A+P_B)\beta - P_B \gamma}{P_B (P_A + P_B)} y_B^t} \right]_{y_B^t = -P_B/2}^{y_B^t = +P_B/2} \right)\\
    =& 4 \dfrac{\sin\left( \pi \frac{(P_A+P_B)\alpha - P_A \gamma}{P_A + P_B} \right)}{2\pi \frac{(P_A+P_B)\alpha - P_A \gamma}{P_A (P_A + P_B)}} \dfrac{\sin\left( \pi \frac{(P_A+P_B)\beta - P_B \gamma}{P_A + P_B} \right)}{2\pi \frac{(P_A+P_B)\beta - P_B \gamma}{P_B (P_A + P_B)}}.
\end{align*}

\section{Calculation of integral for outgoing messages}
\label{app:integral-hypergeometric}
Below is shown the detailed calculation of the integral in \cref{eq:integral-hypergeometric}.
\begin{align}
    I_{\gamma} (x_i^{t+1}, x_j^t) \coloneqq& \int_{-P/2}^{+P/2} \diff{y} p(x_i^{t+1} | y, x_i^t, x_j^t)  u_{\gamma}(y) \nonumber\\
    =& \int_{-P/2}^{+P/2} \diff{y} \dfrac{e^{\beta (y + J_{ji}x_j^t + h_i) x_i^{t+1}}}{2 \cosh(\beta (y + J_{ji}x_j^t + h_i))} e^{i \frac{2\pi}{P} \gamma y} \nonumber\\
    =& \left[ \dfrac{e^{i \frac{2\pi}{P} \gamma y}}{\beta (1+x_i^{t+1}) + i \frac{2\pi}{P} \gamma} e^{\beta (y + J_{ji}x_j^t + h_i) (1+x_i^{t+1})} \right. \times \\
    &\times \left. \prescript{}{2}{F}_1 \left( 1, \dfrac{\beta(1+x_i^{t+1}) + i \frac{2\pi}{P} \gamma}{2\beta}, \dfrac{\beta(3+x_i^{t+1}) + i \frac{2\pi}{P} \gamma}{2\beta}, -e^{2\beta(y + J_{ji}x_j^t + h_i)} \right) \right]_{y=-P/2}^{y=+P/2} \nonumber
\end{align}
where $\prescript{}{2}{F}_1(a,b,c,z)$ is the Gaussian hypergeometric function. For $x_i^{t+1}=-1$ and $\gamma=0$ the integral can be evaluated explicitly as
\begin{equation}
    I_{0}(-1, x_j^t) = \left[ y + J_{ji}x_j^t + h_i - \dfrac{1}{2\beta} \log\left(\beta + \beta e^{2\beta(y + J_{ji}x_j^t + h_i)}\right) \right]_{y=-P/2}^{y=+P/2}.
\end{equation}

\section{Population dynamics}
\label{app:popdyn}
Population dynamics is a standard technique used in Statistical Physics \cite{mezardInformationPhysicsComputation2009,mezardBetheLatticeSpin2001}, and particularly within the cavity method. It parametrizes a probability distribution by a population of samples which is then updated in a cavity method scheme. The overall approach enables to study the properties of a disordered system in the thermodynamic limit. In this work, it has been employed to study the behavior of Kinetic Ising on infinite graphs.\\
The method consists of initializing a population of MPBP messages and updating them until the distribution reaches a quasi-fixed point, where they approximately reproduce the statistical properties of the messages of the infinite system. In general, for an infinite graph with some known distributions for the nodes' degrees, couplings and external fields, the update step is carried out as in \cref{alg:popdyn}.
\begin{tcolorbox}[float*=htbp, width=\textwidth]
\setlength{\algoheightrule}{0pt}
\setlength{\algotitleheightrule}{0pt}
\begin{algorithm}[H]
    \caption{Population dynamics}
    \label{alg:popdyn}
    \SetAlgoLined
    \DontPrintSemicolon
    
    \For{it $\in$ $1:$ maxiter}{
        - Consider the ``central" node $i$ and sample a degree $c$ according to the degree distribution \;
        - Sample $c$ couplings $J_{ki}$ and one external field $h_i$ according to their distributions \;
        - Sample $c$ messages $\set{\mu_{n_j}}{j\in 1:c}$ at random from the population \;
        - Using the $\set{\mu_{n_j}}{j\in 1:c}$ as incoming into $i$, compute the outgoing messages $\set{\nu_{j}}{j\in 1:c}$ (\cref{alg:glauber-mpbp-update}) \;
        - Replace $\set{\mu_{n_j}}{j\in 1:c}$ with $\set{\nu_{j}}{j\in 1:c}$ in the population \;
    }
\end{algorithm}
\end{tcolorbox}
\\
In practice, the iteration considers a fictitious ``central" node and uses it to update a randomly sampled set of messages of the population. After some iterations, the population of messages will converge to a stationary distribution that mimics the properties of the system under study in the thermodynamic limit.\\
Once the messages have converged, one can perform additional iterations, storing selected observables and eventually computing statistics on them.

\section{Large deviations}
\label{app:large-deviations}
One of the advantages of Basis-MPBP compared with Monte Carlo or other sampling-based methods is that it can be used to compute the probabilities of very rare events.
For example, considering a Kinetic Ising free dynamics, the magnetization at the final time $\hat{m} = \frac{1}{N} \sum_i x_i^T$ has a distribution that concentrates on the most probable value.
As a result, it is extremely difficult to compute $p(\hat{m})$ for values of $\hat m$ different from the most probable one by means of sampling, because that would require an enormous number of samples, but it can be done with Basis-MPBP.\\
We aim to compute
\begin{equation}
    g(\hat{m}) \coloneqq -\frac{1}{N} \log \left( \sum_{\overline{\bs{x}}} p(\overline{\bs{x}}) \delta_{\hat{m}, \frac1N \sum_i x_i^T} \right)
\end{equation}
so that the probability of observing a magnetization $\hat{m}$ at time $T$ is $p(\hat{m}) = e^{-N g(\hat{m})}$. 
This large deviation law can be estimated from the Bethe free energy and its Legendre transform as follows. The probability of the free dynamics
\begin{equation}
    p(\overline{\bs{x}}) = \prod_{i=1}^N p_i^0(x_i^0)\prod_{t=1}^{T} p_i^t(x_i^t | x_i^{t-1}, \bs{x}_{\del i}^{t-1}) = e^{-H(\overline{\bs{x}})}
\end{equation}
is reweighted at the final time with a variable external field $\hat{h}\in\mathbb{R}$:
\begin{equation}
    p_{\hat{h}}(\overline{\bs{x}}) = \frac{1}{Z_{\hat{h}}} \prod_{i=1}^N  e^{\hat{h} x_i^T}p_i^0(x_i^0) \prod_{t=1}^{T} p_i^t(x_i^t | x_i^{t-1}, \bs{x}_{\del i}^{t-1}) = \frac{1}{Z_{\hat{h}}} e^{-H(\overline{\bs{x}}) +\hat{h}\sum_i x_i^T}.
\end{equation}
MPBP can be used to estimate the magnetization $\hat{m}(\hat{h}) = \sum_{\overline{\bs{x}}} p_{\hat{h}}(\overline{\bs{x}}) (\frac1N \sum_i x_i^T)$ and the Bethe free energy of the model:
\begin{equation}
\begin{split}
    f(\hat{h}) =& -\frac{1}{N} \log Z_{\hat{h}} = -\frac{1}{N} \log \sum_{\overline{\bs{x}}} e^{-H(\overline{\bs{x}}) +\hat{h}\sum_i x_i^T}\\
    =& -\frac{1}{N} \log \sum_{m} \left( \sum_{\overline{\bs{x}}} e^{-H(\overline{\bs{x}})} \delta_{m, \frac{1}{N} \sum_i x_i^T} \right) e^{N\hat{h}m}\\
    =& -\frac{1}{N} \log \sum_{m} p(m) e^{N\hat{h}m}.
\end{split}
\end{equation}
Defining $F(\hat{h}) \coloneqq e^{-N f(\hat{h})}$ we obtain
\begin{equation}
\begin{split}
    F(\hat h) =& \sum_{m} e^{N \hat{h}m} p(m)\label{eq:fh-gm-system}.
\end{split}
\end{equation}
Hence, one can use Basis-MPBP to compute the Bethe free energy $f(\hat{h})$ for different values of $\hat{h}$, then invert the linear system of equations in \cref{eq:fh-gm-system} to retrieve $p(m)$ and $g(m)=-\frac1N \log p(m)$.
We did this for a tree graph of $N=7$ nodes with a time horizon $T=4$.
This example was small enough that we could compare the results with the true value of $g(\hat{m})$, obtained by extensive enumeration. \Cref{fig:tree-ld-gm} shows the agreement between the values obtained by the procedure described above and the exact ones; there is a 1\% root-mean-square error. Note that the matrix of the system \cref{eq:fh-gm-system} is a rectangular Vandermonde matrix, which is well-known to be ill-conditioned. Even with a good approximation of $f(\hat{h})$, one can get large errors in the estimation of $g(\hat{m})$, especially for large $N$.

In the thermodynamic limit $N \to \infty$, fortunately, this numerical inconvenience is not present thanks to the concentration of the measure $p(m)$. Indeed, we compute $f(\hat h)$ and $\hat m(\hat h)$ using Basis-MPBP, 
and obtain \cref{fig:res-ld} as a parametric plot $\hat h \mapsto \left(\hat{m}(\hat{h}), g(\hat{m}(\hat{h}))\right)$ for $g(\hat{m}(\hat{h})) = f(\hat{h}) + \hat{h} \hat{m}(\hat{h})$.

\begin{figure}
    \includegraphics[width=0.8\linewidth]{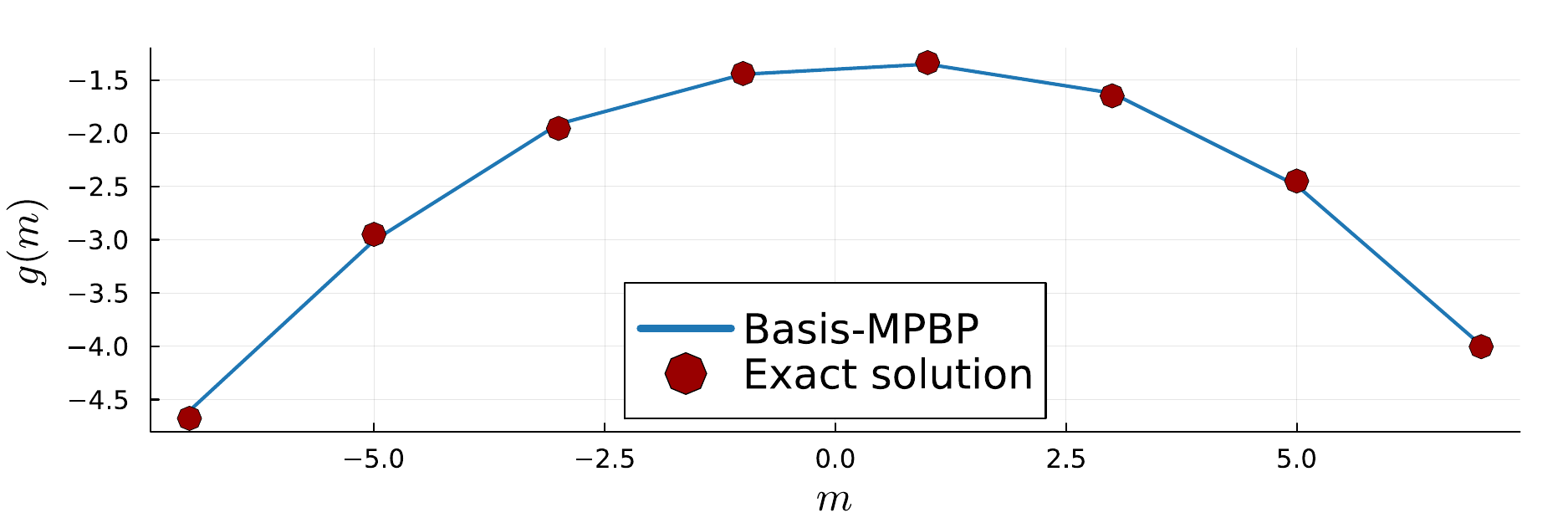}
    \caption{Large deviations law on a random tree with $N=7$ and with $\beta=1/2$, $J\sim \text{Uniform}(0, 1)$, $h=0$, $m^0=0.6$. The algorithm was run with $d=15$ and $K=30$.}
    \label{fig:tree-ld-gm}
\end{figure}

\section{Parameters used in experiments}
\label{app:parameters}
For convenience, we report here the parameters used in every experiment, together with the type of system on which the experiment has been conducted.

\begin{table}[h]
    \centering
    \begin{tabular}{|c|c|c|c|c|c|c|c|c|c|}
        \hline 
        Figure & Graph & $N$ & $T$ & $\beta$ & $m^{0}$ & Distribution of $J_{k,i}$ & $h$ & $d$ & $K$\tabularnewline
        \hline \hline
        \cref{fig:res-tree} & Tree (preferential attachment) & 100 & 10 & 0.5 & 0.7 & Uniform$\left[-1,+1\right]$ & 0 & 15 & 100\tabularnewline
        \hline 
        \cref{fig:res-conf} & Configuration model ($p(c)\sim c^{-0.6}$) & 200 & 10 & 0.7 & 0.6 & Uniform$\left[-1,+1\right]$ & 0 & 20 & 100\tabularnewline
        \hline 
        \cref{fig:res-rr} & Random regular ($c=8$) & $\infty$ & 10 & 0.3 & 0.6 & Uniform$\left[-1,+1\right]$ & 0 & 15 & 60\tabularnewline
        \hline 
        \cref{fig:res-er} & Erdős-Rényi ($\left\langle c\right\rangle =4$) & $\infty$ & 10 & 0.5 & 0.6 & Uniform$\left[-1,+1\right]$ & 0 & 15 & 60\tabularnewline
        \hline 
        \cref{fig:res-ld} & Random regular ($c=8$) & $\infty$ & 10 & 1/7 & 0.6 & Uniform$\left[0,1\right]$ & 0 & 10 & 30\tabularnewline
        \hline 
    \end{tabular}
    \label{tab:parameters}
\end{table}

\end{onecolumngrid}

\bibliography{references}
\clearpage

\end{document}